\documentclass[pre,twocolumn]{revtex4}
\usepackage[normalem]{ulem}
\usepackage{amsmath,amssymb}
\usepackage[ps2pdf,bookmarks=true,colorlinks,linkcolor=red,urlcolor=blue,citecolor=blue]{hyperref}
\usepackage{graphicx,amsmath,amsfonts}
\usepackage[usenames]{color}
\usepackage{epstopdf}

\begin{document}

\title{Universal dynamical phase diagram of lattice spin models and
strongly correlated ultracold atoms in optical lattices. }

\author{E.A. Demler $^{1}$}
\author{A.Ya. Maltsev $^{2}$}
\author{A.O. Prokofiev $^{2}$}

\affiliation{$^{1}$ Physics Department, Harvard University,
Cambridge, MA 02138, USA}
\affiliation{$^{2}$ L.D.Landau Institute for Theoretical Physics
142432 Chernogolovka, Moscow reg., Russia}

\date{\today}

\begin{abstract}
We study semiclassical dynamics of anisotropic Heisenberg models in
two and three dimensions. Such models describe lattice spin systems
and hard core bosons in optical lattices. We solve numerically
Landau-Lifshitz type equations on a lattice and show that in the phase
diagram of magnetization and interaction anisotropy, one can identify
several distinct regimes of dynamics. These regions can be distinguished
based on the character of one dimensional solitonic excitations, and
stability of such solitons to transverse modulation. Small amplitude
and long wavelength perturbations can be analyzed analytically using mapping of non-linear
hydrodynamic equations to KdV type equations. Numerically we find that properties
of solitons and dynamics in general remain similar to our analytical results
even for
large amplitude and short distance inhomogeneities, which allows us to obtain
a universal dynamical phase diagram. As a concrete example we study dynamical
evolution of the system starting from
a state with magnetization step and show that formation of oscillatory
regions and their stability to transverse modulation can be understood
from the properties of solitons. In regimes unstable to transverse
modulation we observe formation of lump type solutions with modulation
in all directions.  We discuss implications of our results for
experiments with ultracold atoms.
\end{abstract}

\pacs{xx.xx}

\maketitle

\begin{figure}[t]
\begin{center}
\includegraphics[width=1\linewidth]{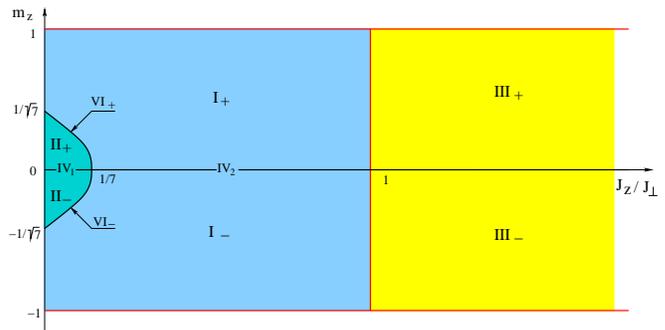}
\end{center}
\vspace{-.3in}
\caption{
Dynamical phase diagram of the anisotropic Heisenberg model
(\ref{heisenberg}) for initial states with finite magnetization
in the XY plane. Vertical axis labels  magnetization
for spin models and density for hard-core bosons on a lattice.
Regions I, II correspond to easy plane anisotropy of the Hamiltonian,
linearized collective modes are of the hyperbolic type. Regions III
correspond to easy axis anisotropy, linearized collective modes
are of the elliptic type.
Special lines $J_z/J_\perp=1$ and $m_z = \pm 1$ all have linearized collective modes
of parabolic type.
Region I$_+$ has particle solitons unstable to transverse modulation.
Region I$_-$ has hole solitons unstable to transverse modulation.
Region II$_+$ has hole solitons stable to transverse modulation.
Region I$_-$ has particle solitons stable to transverse modulation.
}
\label{PhaseDiagram}
\end{figure}

{\it Motivation}.
Understanding nonequilbrium quantum dynamics of many-body systems
is an important problem in many areas of physics. The most fundamental
challenge facing this field is identifying emergent collective phenomena,
which should lead to universality and common properties even in systems
which do not have identical microscopic Hamiltonians. We know that in
equilibrium many-body systems typically fall within certain universality
classes. Basic examples are  states with spontaneously broken symmetries
such as BEC of bosons and superfluid paired phases of fermions\cite{Giorgini2008,Dalfovo1999}, Fermi
liquid states of electrons and interacting ultracold fermions\cite{Shankar1994,Nascimbene2010}. We have
numerous examples of emergent universal properties of classical systems
driven out of equilibrium (see ref. \cite{Cross1993} for a review).
However very few examples of collective behavior of coherent quantum evolution of many-body
systems, which can be classified as exhibiting emergent universality, are known
(several recently studied examples can be found in refs \cite{Kinoshita2006,Moeckel2008,Rigol2008,Barmettler2009,Polkovnikov2011,Schneider,Gring2011})
The main result of this paper is  demonstration of the universality
of semiclassical dynamics of lattice spin systems and strongly interacting
bosons in two and three dimensions, summarized in the phase diagram in Fig.
\ref{PhaseDiagram}.

In addition to fundamental conceptual importance, our study has strong
experimental motivation. Rapid progress of experiments with ultracold
atoms in optical lattices makes it possible to create well controlled
realizations of systems that we discuss \cite{Bloch2008,Lewenstein2007a}.
Tunability of such systems,
nearly perfect isolation from the environment, and a rich toolbox of
experimental probes, including the possibility of single site resolution \cite{Nelson2007,Gemelke2009,Sherson2010a,Bakr2010a},
makes them excellent candidates for exploring nonequilibrium quantum
dynamics (see \cite{Polkovnikov2011,Lamacrafta} and references therein).
Several intriguing nonequilibrium phenomena have been
demonstrated recently in ensembles of ultracold atoms, including
collapse and revival of coherence\cite{Bloch2002}, absence of relaxation in nearly
integrable systems\cite{Kinoshita2006}, exponential slowdown of relaxation in systems with strong mismatch of excitation energies\cite{Strohmaier2010}, anomalous diffusion\cite{Schneider},
prethermalization\cite{Gring2011}, many-body Landau-Zener transitions \cite{Chen2010},
light-cone spreading of correlations in a many-body system\cite{Cheneau2011}.

{\it Model}. In this paper we study non-equilibrium
dynamics of lattice spin models and strongly correlated bosons in optical
lattices. Our starting point is the anisotropic Heisenberg model
\begin{eqnarray}
{\cal H}_{\rm AH} = - J_{\perp} \sum_{\langle ij \rangle}
\left( \sigma_{i}^{x} \sigma_{j}^{x} + \sigma_{i}^{y}
\sigma_{j}^{y} \right)
- J_{z} \sum_{\langle ij \rangle} \sigma_{i}^{z} \sigma_{j}^{z}
\label{heisenberg}
\end{eqnarray}
where $\sigma^{a}$ are Pauli matrices.
Anisotropic Heisenberg models with tunable interactions
can be realized with two component
Bose mixtures in spin dependent optical lattices\cite{Duan2003,Kuklov2003,Altman2003,Trotzky2008}.
Hamiltonian (\ref{heisenberg}) also describes spinless bosons
in the regime of infinitely strong on-site repulsion\cite{Batrouni1995}.
In this case states
$| \downarrow \rangle$ and $| \uparrow \rangle$ correspond to states
with zero and one boson per site respectively,
$J_{\perp}$ is the tunneling strength, and $J_z$ is the strength of the nearest
neighbor repulsion. For atoms with contact interaction and
confined to the lowest Bloch band $J_z=0$.  Non-local interactions
can be present for atoms
in higher Bloch bands\cite{Scarola2005} and polar molecules\cite{Lahaye2009}.

Semiclassical equations of motion can be understood either as the
lattice version of Landau-Lifshitz equations or as dynamics in the
manifold of variational states
$| \Psi(t) \rangle = \prod_{i} [ \sin {\theta_{i} \over 2}
e^{-i \varphi_{i}/2} |\downarrow \rangle_{i} +
\cos {\theta_{i} \over 2}  e^{i \varphi_{i}/2}
|\uparrow \rangle_{i} ]$,
where all $\theta_{i}$ and $ \varphi_{i}$ are independent functions
of time ( see supplementary material and ref. \cite{Demler2011a}).
It is convenient to introduce parametrization

$
\sin {\theta_{i} \over 2}  =  ({1 - z_{i} \over 2})^{1/2}$,
$\cos  {\theta_{i} \over 2}  =  ({1 + z_{i} \over 2})^{1/2}$ ,
$e^{i\varphi_{i}}  =
{x_{i} + i y_{i} \over {(x_{i}^{2} + y_{i}^{2}})^{1/2}}
$,
then semiclassical dynamical equations are given by

\begin{equation}
\label{xyzSystem}
\begin{array}{c}
{\dot x_{i}} \,\, = \,\, 4 \, J_{\perp} \, z_{i} \,
\left( y_{i+1} + y_{i-1} \right) \, - \, 4 \, J_{z} \, y_{i} \,
\left( z_{i+1} + z_{i-1} \right)
\cr
{\dot y_{i}} \,\, = \,\, - \, 4 \, J_{\perp} \, z_{i} \,
\left( x_{i+1} + x_{i-1} \right) \, + \, 4 \, J_{z} \, x_{i} \,
\left( z_{i+1} + z_{i-1} \right)
\cr
{\dot z_{i}} \, = \, 4 \, J_{\perp} \, \left(
y_{i} \, x_{i+1} \, - \, x_{i} \, y_{i+1} \, + \,
y_{i} \, x_{i-1} \, - \, x_{i} \, y_{i-1} \right)
\end{array}
\end{equation}
Note that variables $(x_{i}, y_{i}, z_{i})$ reside on a sphere.
Specifying $x_{i}^{2} \, + \, y_{i}^{2} \, + \, z_{i}^{2} \, = 1$
in the beginning of the evolution will preserve this condition
at all times.

We consider dynamics in a state with finite magnetization in the $XY$ plane
(superfluid phase  for hard core bosons), where close to equilibrium dynamics is determined by
the Bogoliubov (Goldstone) mode. Considerable difference
between the nonlinear hydrodynamics of hard core bosons on a lattice
and the more familiar GP model was observed  in the numerical
study of solitary waves in ref. \cite{Balakrishnan2009}(for a discussion of
solitons in systems of ultracold atoms in the regime where GP equation applies see refs. \cite{Burger1999,Johansson1999,Denschlag2000,Trombettoni2001,Strecker2002,Khaykovich2002,Baizakov2003,Kevrekidis2003,Ahufinger2004,Eiermann2004,Becker2008,Scott2011}). Additional evidence for the special character of solitons in systems of strongly
correlated lattice bosons came from the analytical study in Ref. \cite{Demler2011a}, which extended
linear hydrodynamics to include the first non-linear term and the first
dispersion correction to the long wavelength expansion and showed that solitons
were of the KdV type. In this paper we derive a full phase diagram of semiclassical
dynamics of (\ref{heisenberg}) for initial states with finite magnetization in the XY
plane. Our approach relies on combining analytical results
describing small density perturbations \cite{Demler2011a} with
numerical studies of equations (\ref{xyzSystem}) describing
dynamics of large amplitude modulations.


{\it Phase diagram. Hyperbolic, elliptic, and parabolic regimes}. The main  results of our analysis are summarized in fig. \ref{PhaseDiagram}.
Firstly  there is fundamental difference between elliptic and hyperbolic regions
in the character of linearized equations of motion.
The hyperbolic regime (hyperbolic regions I and II in fig. \ref{PhaseDiagram})
corresponds to the easy plane anisotropy of the Hamiltonian. If we start with initial
state that has finite magnetization in the XY plane, we find collective modes with linear dispersion, $\omega \propto | q |$, which describe
Bogoliubov (Goldstone) modes. In the  elliptic regime we  have easy axis anisotropy:
lowest energy configuration should have full polarization along the z-axis, $m = \pm 1$. However our initial state
has magnetic order in the XY plane. Thus we find unstable collective modes with imaginary frequency
${\it Im}\,  \omega \propto  \pm | q |$. This corresponds to the dynamical instability in which
small density fluctuations grow in time exponentially at short time scales.
There are also several special lines on the phase diagram \ref{PhaseDiagram}:
the SO(3) symmetric Heisenberg model with
$J_z=J_\perp$ and  fully polarized regimes $m = \pm 1$
(fully occupied or empty regimes for spinless bosons). In all of these cases collective modes have
quadratic dispersion, i.e. linearized equations of motion are of the
parabolic type.


{\it Hyperbolic regime. Solitons}. Hyperbolic regions I and II are differentiated further according
to the character of solitonic excitations. In region I$_+$ we find
particle-like one dimensional solitons, which are
unstable to modulation in the transverse directions. In region I$_-$ we
find hole-like one dimensional solitons  unstable to transverse
modulations. In regions II$_{\pm}$ we find particle and hole-like
solitons stable to transverse modulation.
Lines ${\rm IV}_{1}$ and
${\rm IV}_{2}$ correspond to two
different regimes of the mKdV equation. Both hole and
particle-type soliton solutions can be found on the
line ${\rm IV}_{2}$. Both types of solitons are
suppressed close to the line ${\rm IV}_{1}$.

In the limit of small amplitude and long wavelength inhomogeneities
non-linear hydrodynamic equations (\ref{xyzSystem}) can be
mapped to KdV type equations and solitons can be
studied analytically (see supplementary material and ref. \cite{Demler2011a}).
This analysis can be used to obtain
explicit expressions for boundaries of different regimes
in fig. \ref{PhaseDiagram}. In the general case
of large amplitude inhomogeneities one needs to solve
lattice equations (\ref{xyzSystem}) numerically. Surprisingly we find
that in even for large amplitude solitons, their properties, including stability to transverse
fluctuations, are consistent with small amplitude solitons. Hence the phase diagram in fig.
\ref{PhaseDiagram} is generic.

\begin{figure}

\begin{tabular}{cc}
\includegraphics[width=0.5\linewidth]{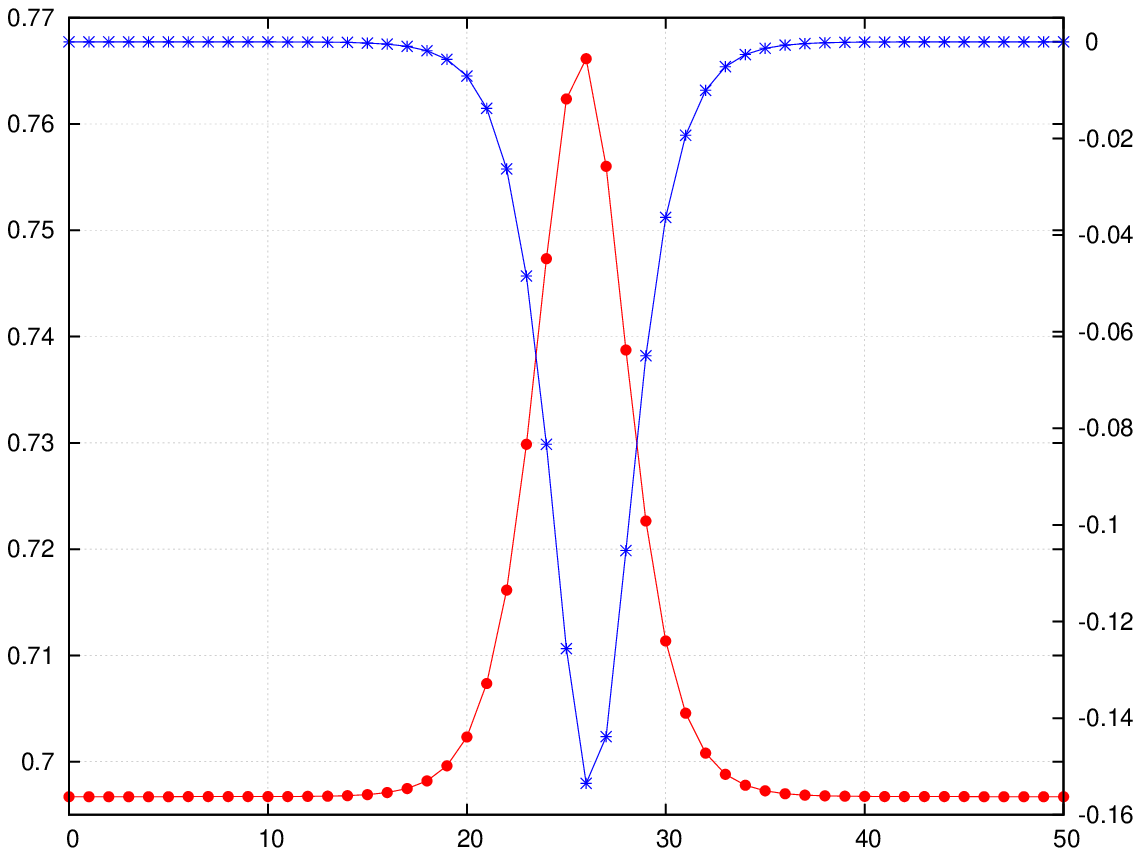} &
\includegraphics[width=0.5\linewidth]{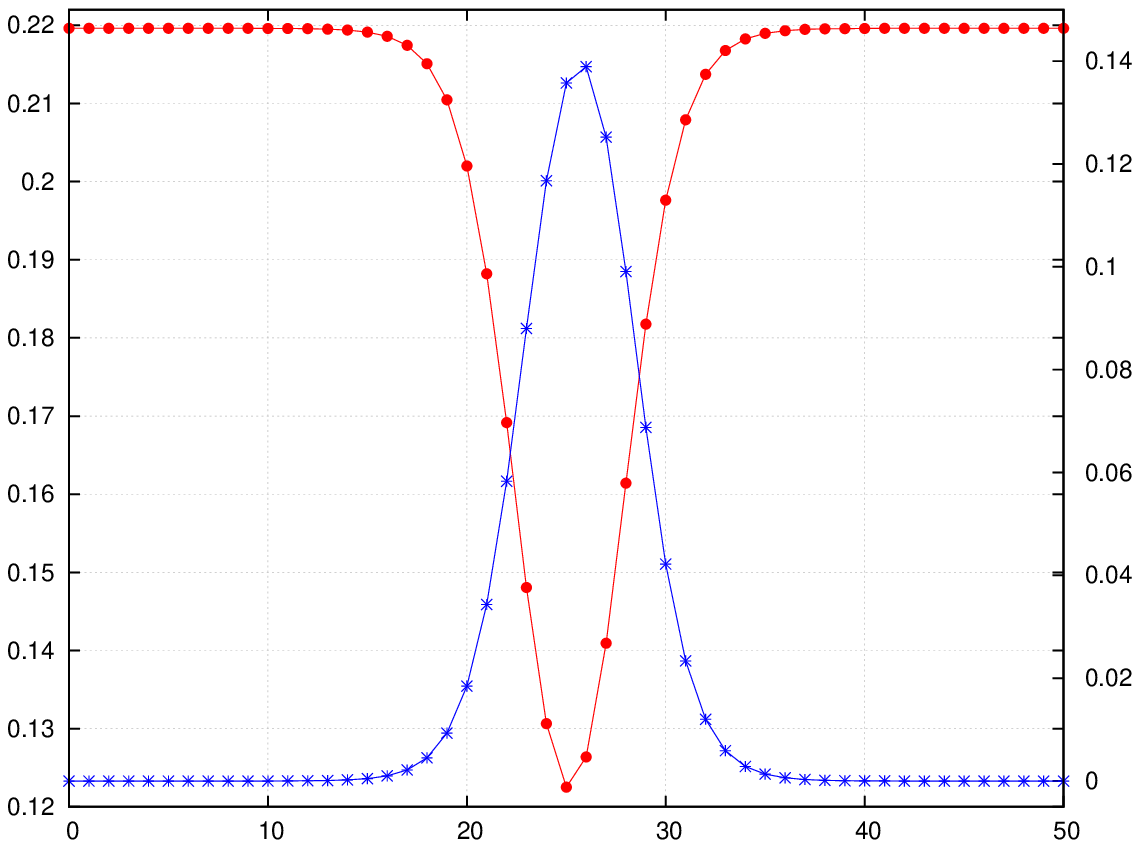}
\end{tabular}
\caption{Left figure: particle soliton in region I$_{+}$. Right figure:
hole soliton in region II$_{+}$. Red line shows magnetization
(density) $\cos \theta_i$. Blue line shows superfluid velocity
$\varphi_{i+1} - \varphi_{i}$. Results obtained
from numerical solutions of lattice equations (\ref{xyzSystem}).
Transverse modulation was not included.
}
\label{SolitonPlusMinus}

\end{figure}

\begin{figure}

\begin{tabular}{cc}
\includegraphics[width=0.5\linewidth]{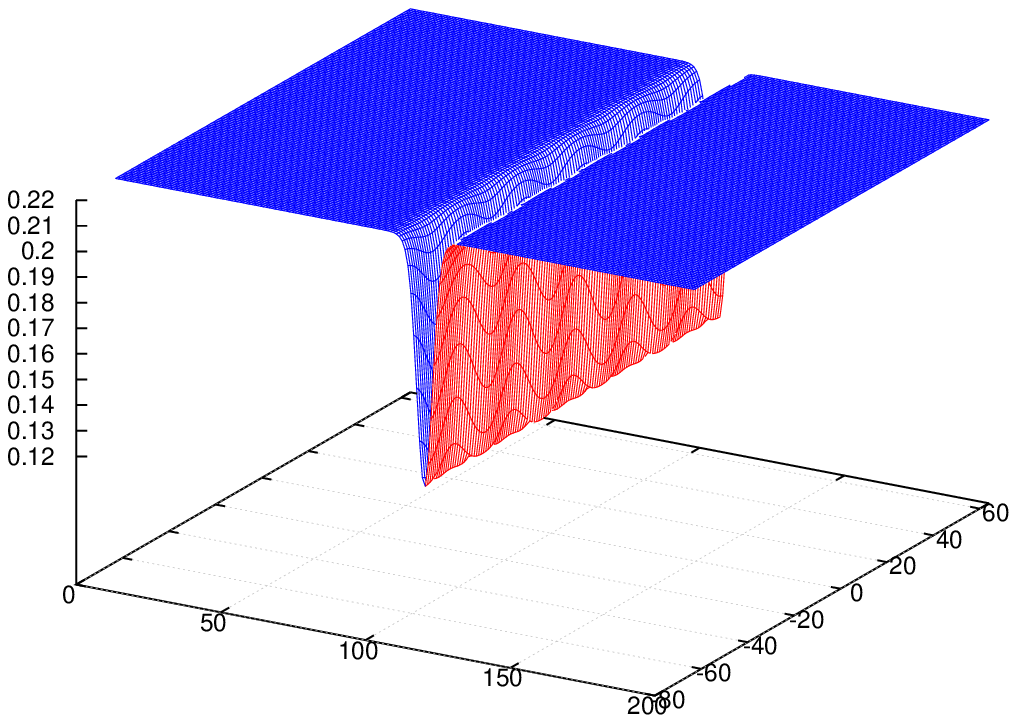} &
\includegraphics[width=0.5\linewidth]{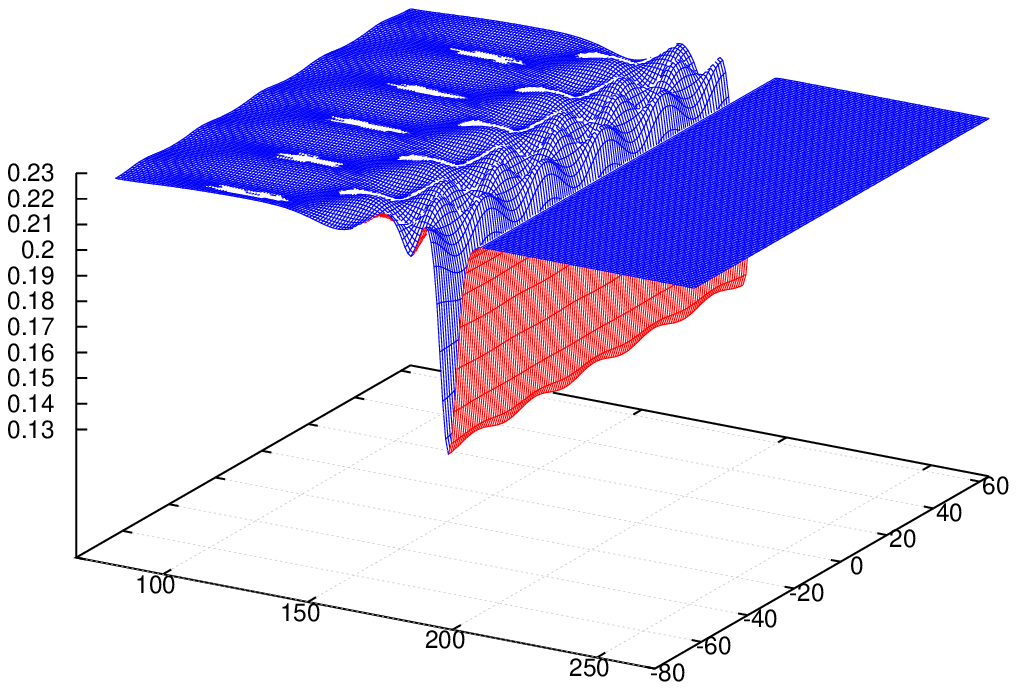} \cr
\includegraphics[width=0.5\linewidth]{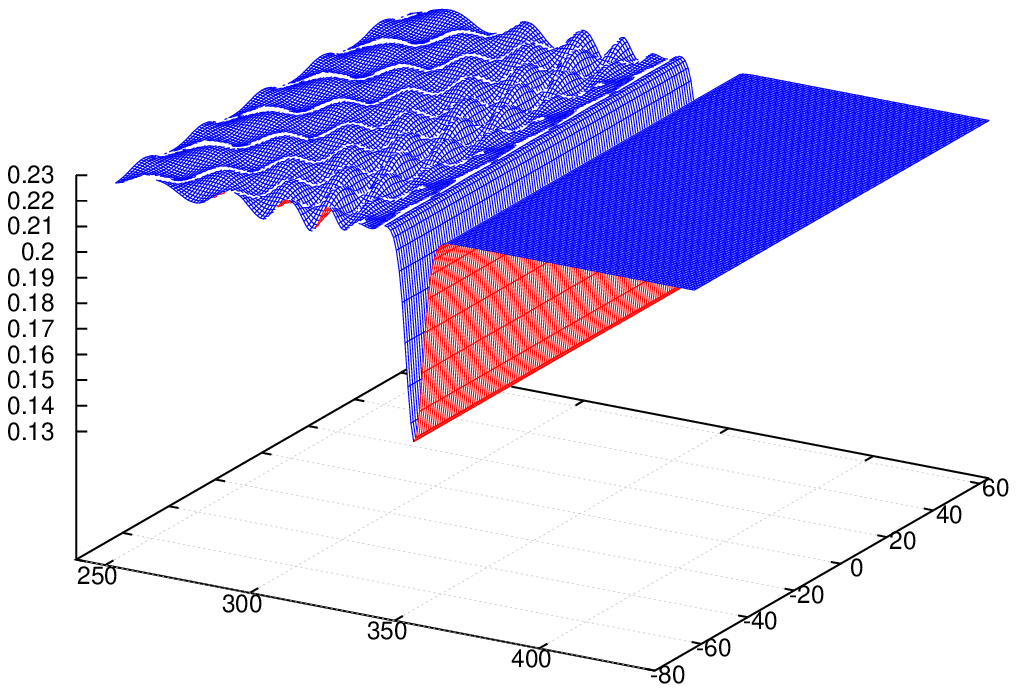} &
\includegraphics[width=0.5\linewidth]{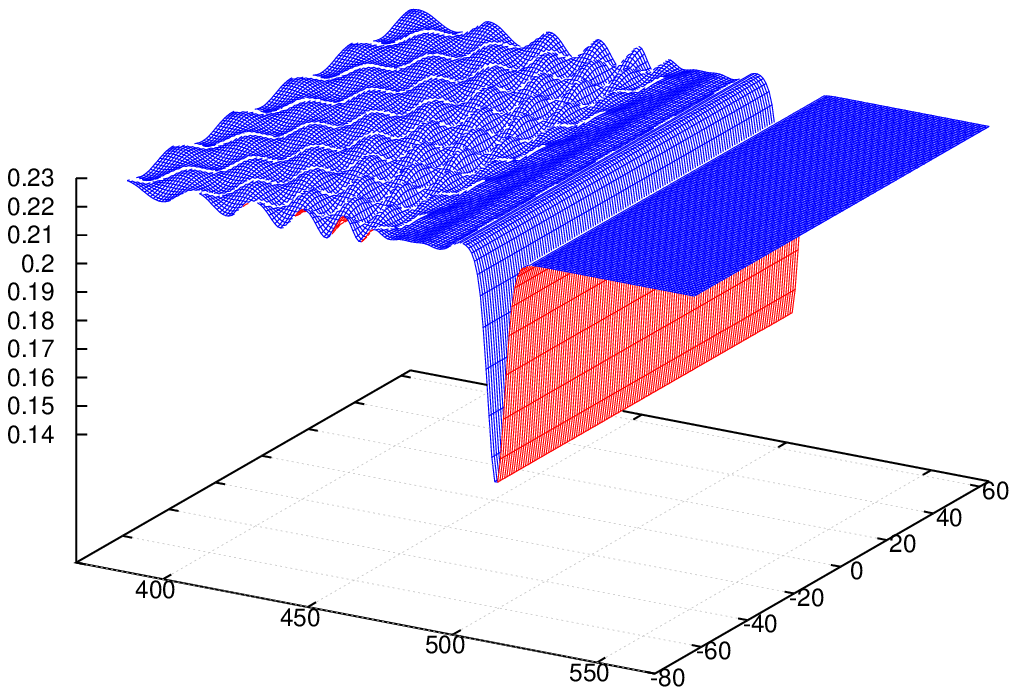}
\end{tabular}
\caption{Time evolution of a soliton for $J_z/J_\perp = 0$ in
region II$_{+}$ in a 2d system. Horizontal axis are spatial coordinates.
Vertical axis is magnetization (density).
Different plots correspond to $t J_\perp = 0$, $10$, $30$,
$50$. This soliton is stable to to small transverse modulation
introduced explicitly in the initial state. }
\label{TwoDimMinus}

\end{figure}

\begin{figure}

\begin{tabular}{cc}
\includegraphics[width=0.5\linewidth]{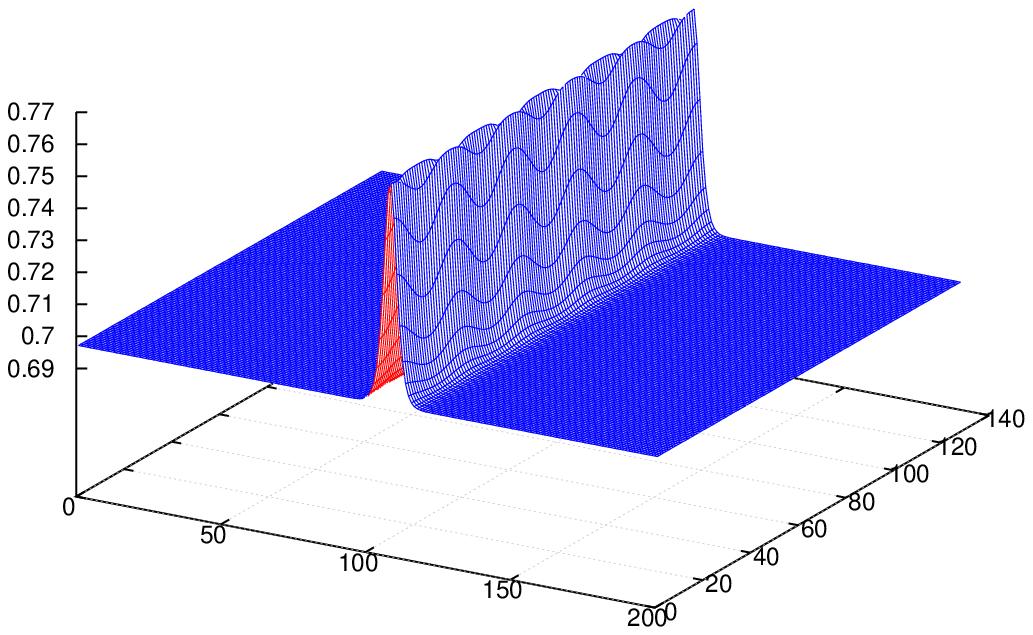} &
\includegraphics[width=0.5\linewidth]{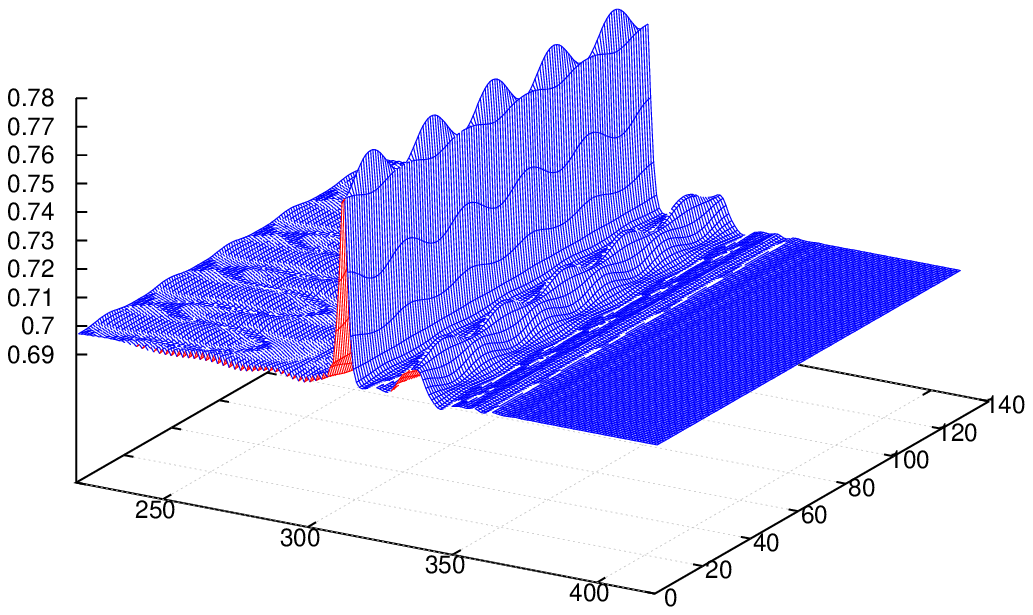} \cr
\includegraphics[width=0.5\linewidth]{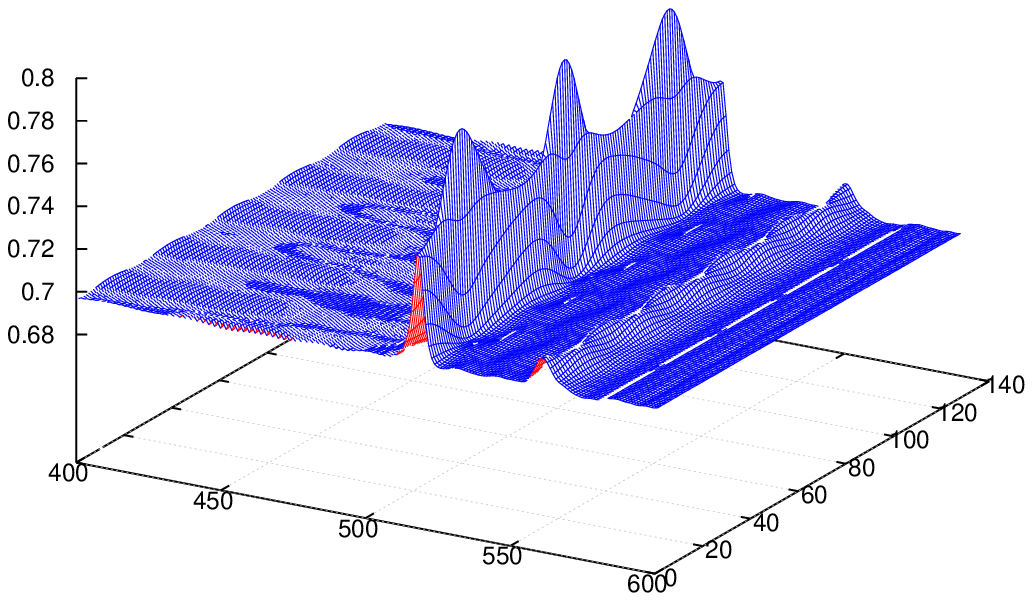} &
\includegraphics[width=0.5\linewidth]{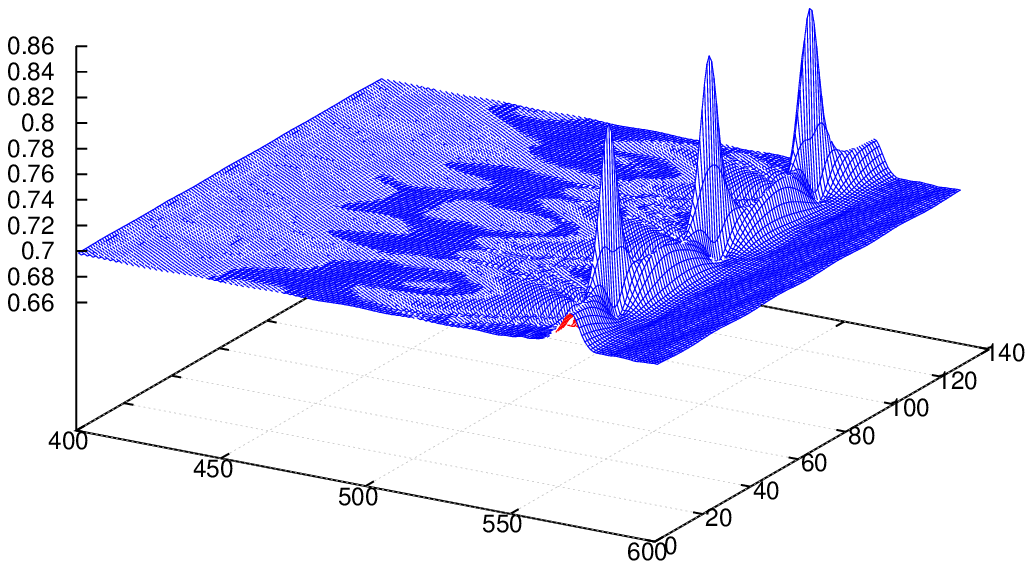}
\end{tabular}
\caption{Time evolution of a soliton for $J_z/J_\perp = 0$ in
region I$_{+}$ in a 2d system. Axis are the same as in fig. \ref{TwoDimMinus}
Different plots correspond to $t J_\perp = 0$, $40$, $80$,
$90$. This soliton is {\it unstable} to small transverse modulation
introduced explicitly in the initial state.
}
\label{TwoDimPlus}

\end{figure}

While we can not present all of our numerical results, we provide
examples, which demonstrate properties of large amplitude solitons.
Fig.  \ref{SolitonPlusMinus} shows  particle- and hole-type solitons in
regions I$_{+}$ and II$_{+}$ respectively.  These specific solutions
were chosen arbitrarily from a plethora of solitons with different amplitudes
and velocities, which can be found in the system for the same values of
microscopic parameters. In these specific examples
only one dimensional variation of parameters were considered
when solving lattice equations (\ref{xyzSystem}). The possibility of transverse modulation
of one dimensional solitons in two dimensional systems
is considered in figures  \ref{TwoDimMinus}, \ref{TwoDimPlus}.
These figures contrast  stable and unstable regimes. In the stable regime II$_+$
transverse modulation does not lead to any dramatic change of the
soliton solution. Analogous dynamics takes place in the other stable regime, II$_-$,
except for a change from hole solitons to particle ones. In the unstable regimes I$_{+}$ (the same
behavior is observed in I$_-$) we observe that transverse modulation
leads to the formation of two-dimensional "lump" solutions.
For small amplitude modulations stability to transverse modulation
can be studied analytically using Kadomtsev-Petviashvili equation \cite{Kamchatnov2008,Demler2011a}.
Our analysis of the stability  to transverse modulation is not limited to
solitons but applies to all inhomogeneous states.

{\it Parabolic and elliptic  regimes}.
A rapidly growing oscillation zone characterizes
elliptic instability  for system (\ref{xyzSystem}) in region III on the phase
diagram. It is shown in fig. \ref{Elliptic}.
In supplementary material we  show solution
of (\ref{xyzSystem}) in the parabolic regime separating
the elliptic and hyperbolic regions on the phase diagram
\ref{PhaseDiagram}.

\begin{figure}

\begin{tabular}{cc}
\includegraphics[width=0.45\linewidth]{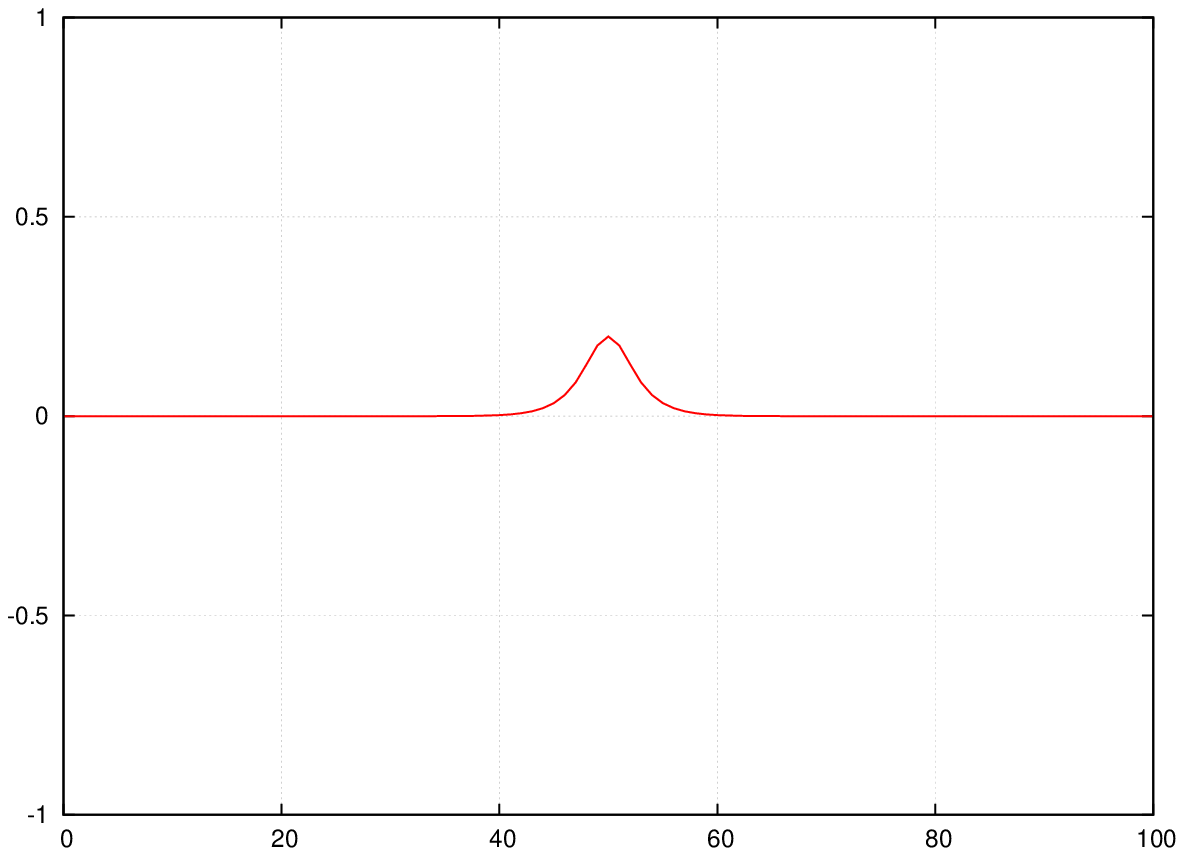} &
\includegraphics[width=0.45\linewidth]{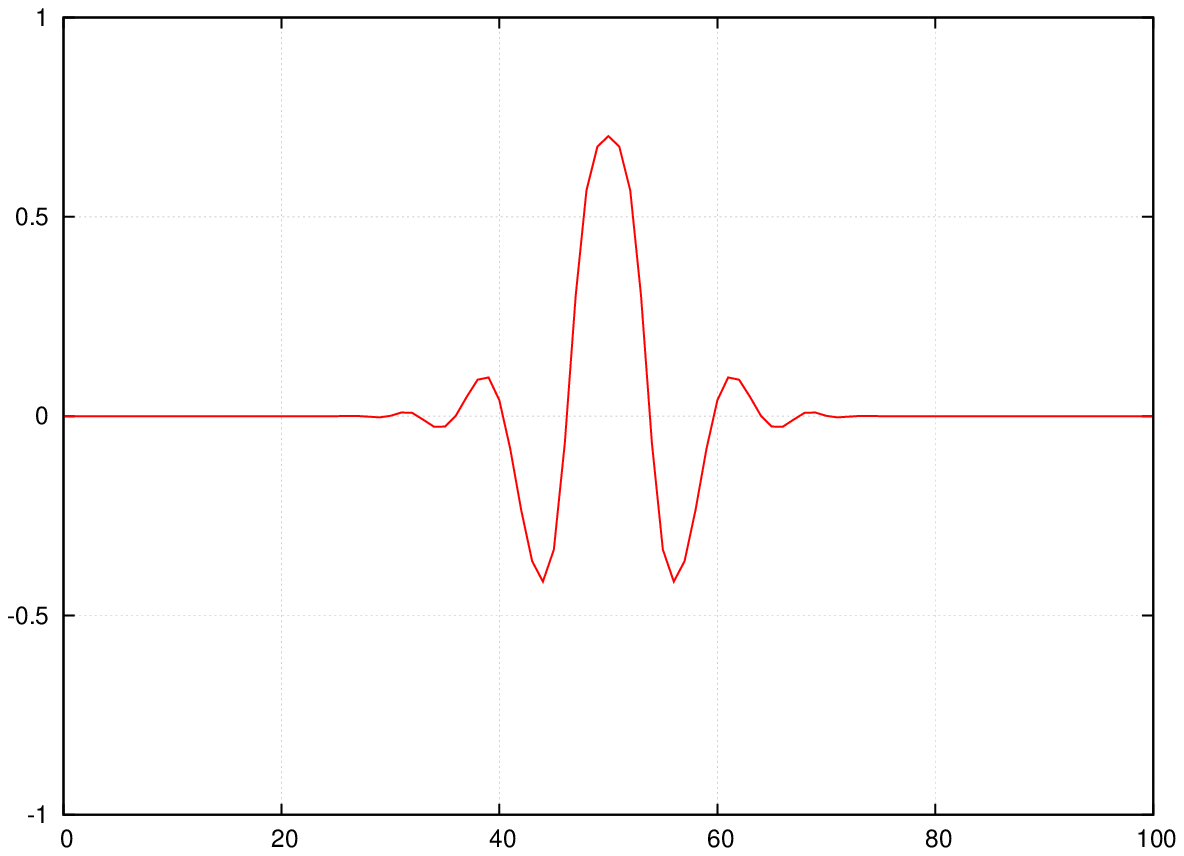} \cr
\includegraphics[width=0.45\linewidth]{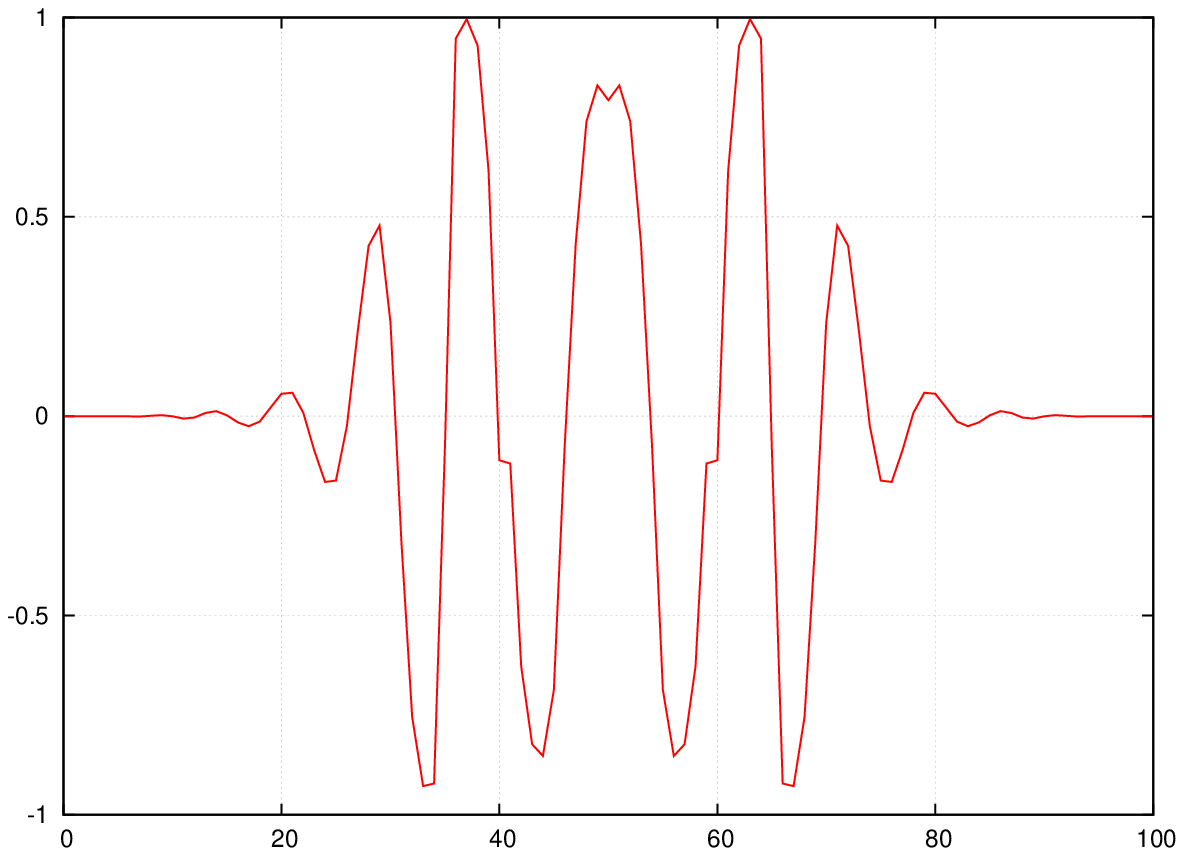} &
\includegraphics[width=0.45\linewidth]{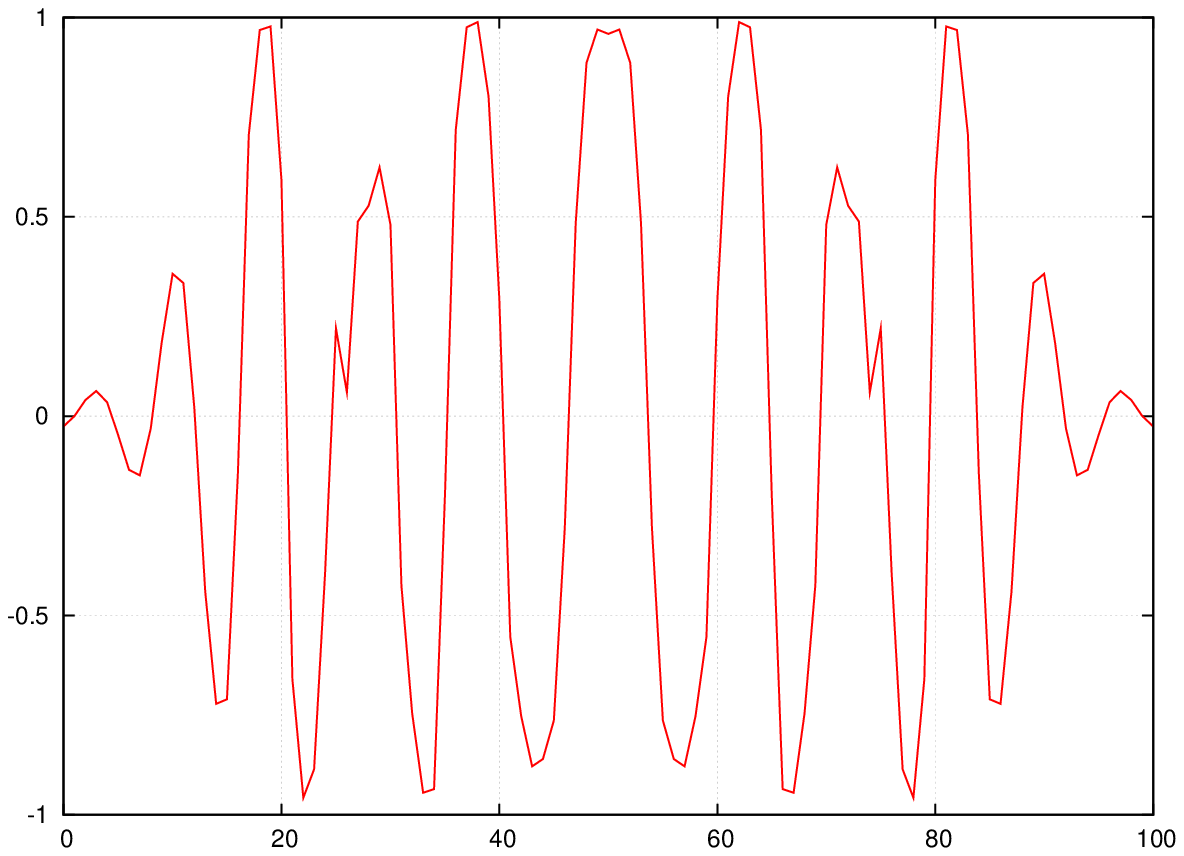}
\end{tabular}
\caption{Time evolution of a small density
perturbation in the elliptic region. Vertical axis is
magnetization (density).
Note rapidly developing oscillation zone typical
to modulation instability.
Different plots correspond to $t J_\perp = 0$, $2$, $4$,
$6$.}
\label{Elliptic}

\end{figure}

{\it Decay of magnetization step}. We now discuss the
problem of the decay of magnetization step in a system of type
(\ref{heisenberg}). Configurations of this type have been recently
realized in 3d magnetic systems using tunable magnetic field gradient\cite{Medley2011}.
They can  be created for strongly correlated spinless
bosons in 2d optical lattices using local addressability \cite{Bakr2010a,Sherson2010a}.
Besides this experimental motivation, relaxation of the
magnetization (density) step is an important methodological problem.
Dynamics starting from this state allows extended unattenuated propagation of the
magnetization (density) modulation, which should amplify the role of nonlinearities. This may be contrasted
to e.g. dynamics starting with a ring type inhomogeneity, where
already at the level of linear hydrodynamics there is a decrease
of the magnetization (density) modulation in time.
In classical hydrodynamics decay of the density step is one of the canonical problems
considered in Refs. \cite{Gurevich1973,Gurevich1974,Lax1979}.

\begin{figure}

\begin{tabular}{cc}
\includegraphics[width=0.5\linewidth]{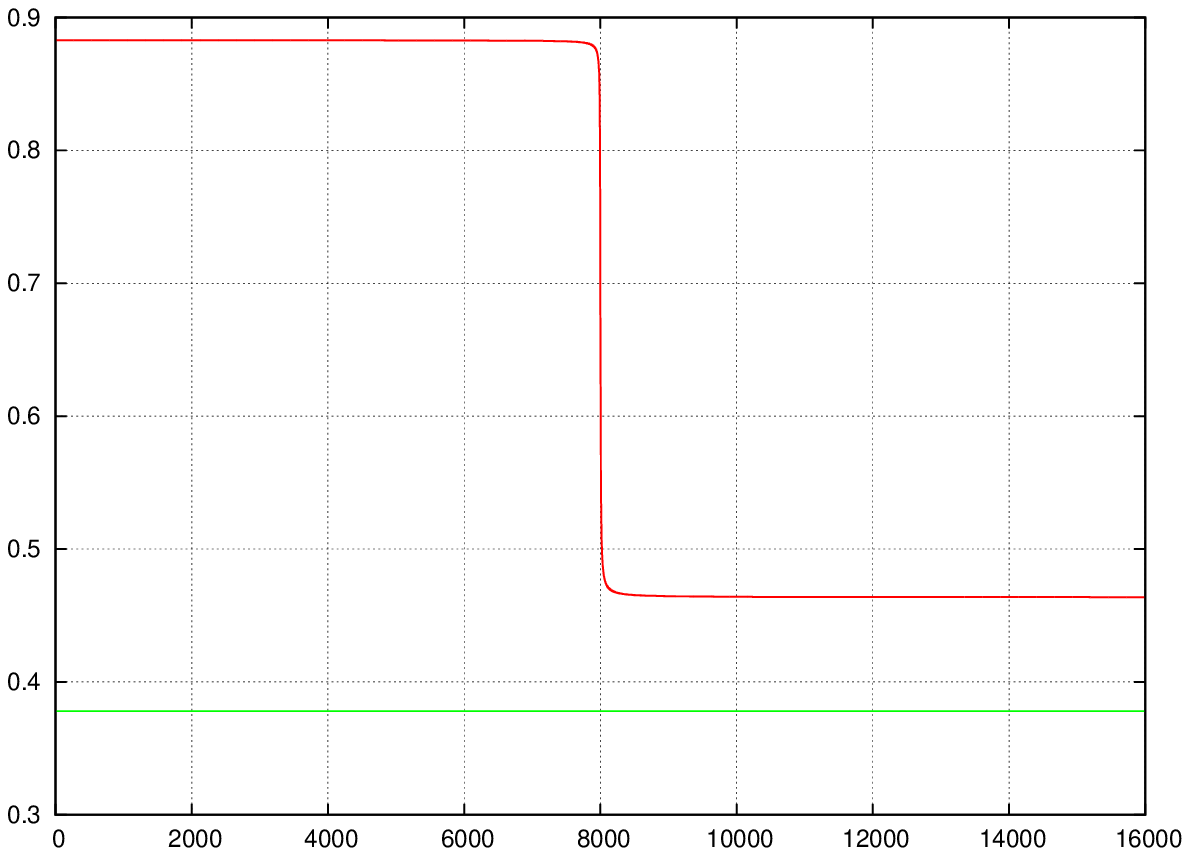} &
\includegraphics[width=0.5\linewidth]{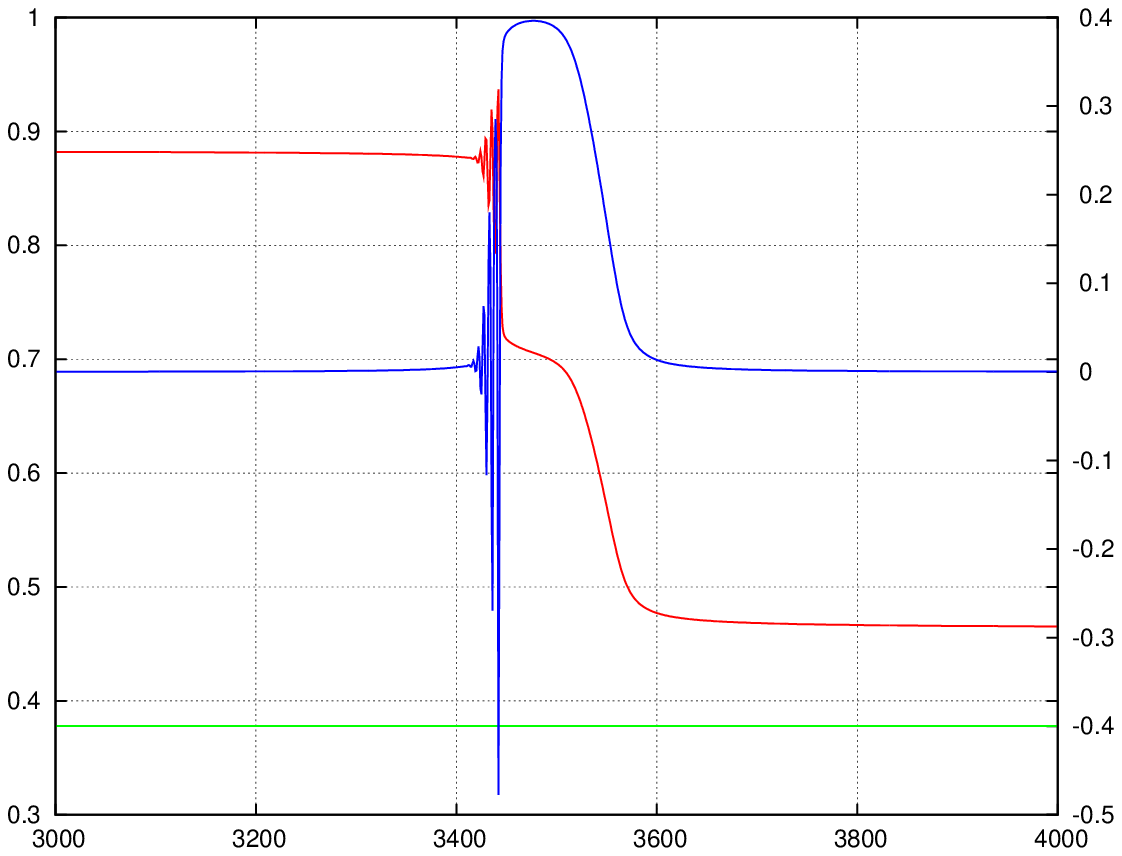} \cr
\includegraphics[width=0.5\linewidth]{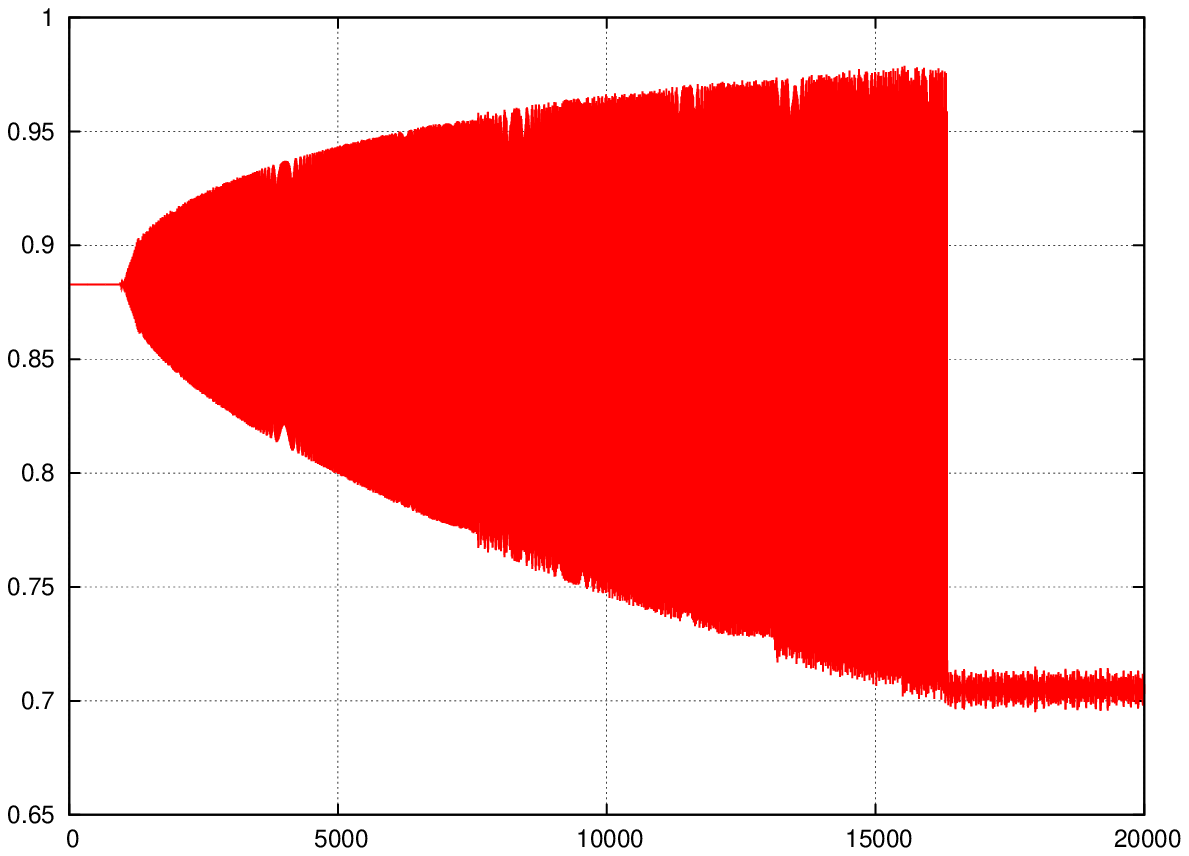} &
\includegraphics[width=0.5\linewidth]{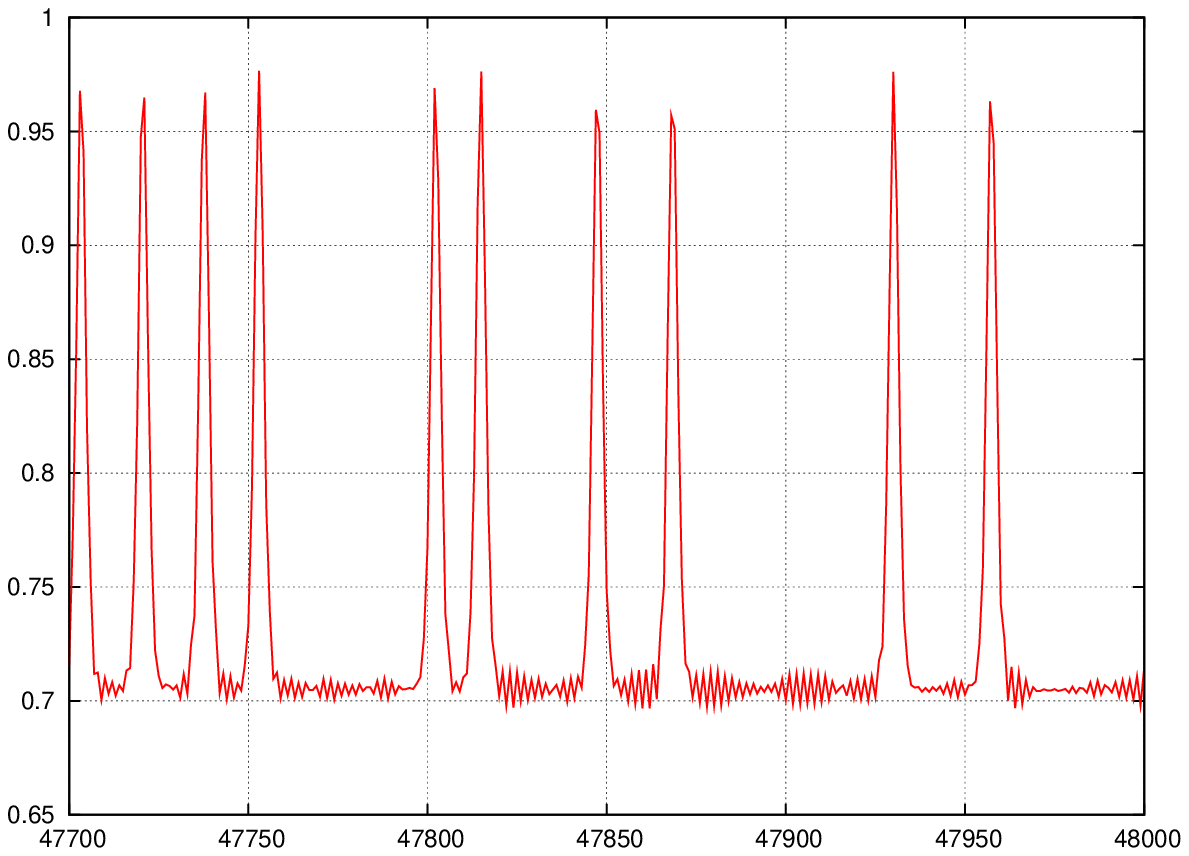}
\end{tabular}
\caption{Decay of the density step in Region I$_{+}$. Red lines
describe density, blue line corresponds to the phase difference.
(a) Initial state $t=0$. (b) After $tJ_\perp=25$ of the evolution. Note separation
of the right and left moving parts and formation of the oscillatory
region on the left moving part.  (c) Expanded view of the left front
after $tJ_\perp=5000$ showing
formation of the leading edge of the oscillation zone.
(d) Left front after $tJ_\perp=15000$. Note soliton pairing mediated by background waves with
small amplitude.}
\label{SolitInter}

\end{figure}

Fig. \ref{SolitInter} shows the main stages of the magnetization step
decay in regions I$_\pm$, II$_\pm$, IV: separation of left- and right-moving parts,
steepening of one of the moving edges and formation of the oscillatory front. While there
is general agreement between small amplitude limit analyzed in ref. \cite{Demler2011a}
and results of numerical analysis of lattice equations
presented in fig. \ref{SolitInter}, there is one important difference.
In the oscillatory region appearing from the decay of a large amplitude magnetization step we observe pairing of solitons,
which can be understood as a result of interaction
between solitons through the background of small amplitude waves.
This is manifestation of the absence of exact integrability of
lattice equations (\ref{xyzSystem}).

\begin{figure}

\begin{tabular}{cc}
\includegraphics[width=0.5\linewidth]{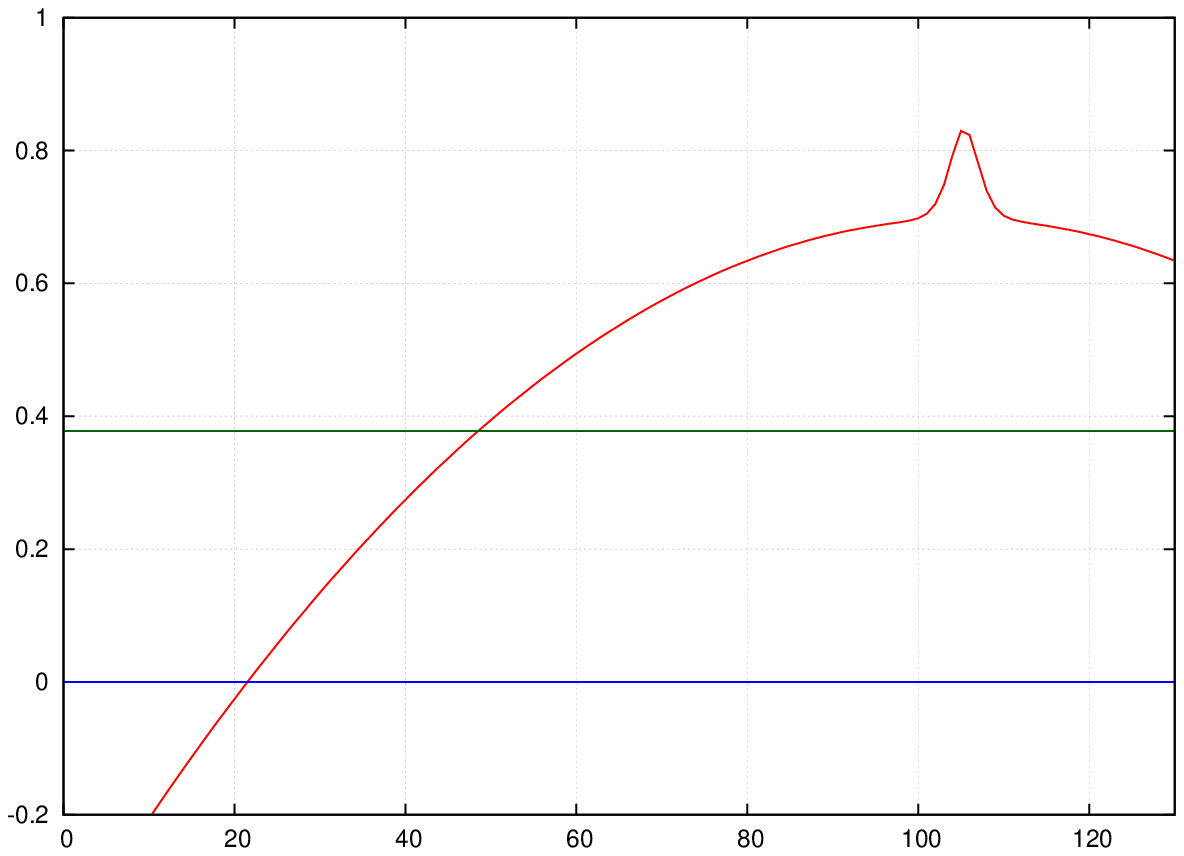} &
\includegraphics[width=0.5\linewidth]{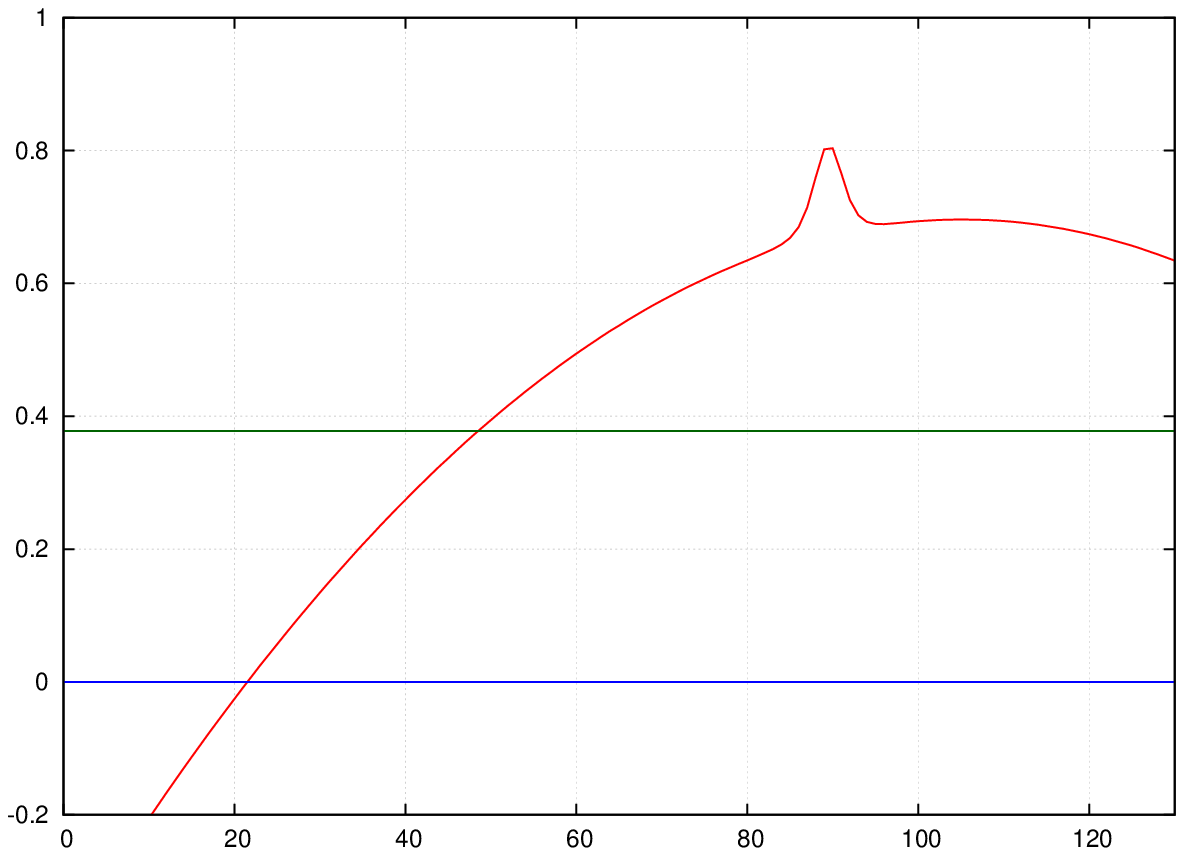} \cr
\includegraphics[width=0.5\linewidth]{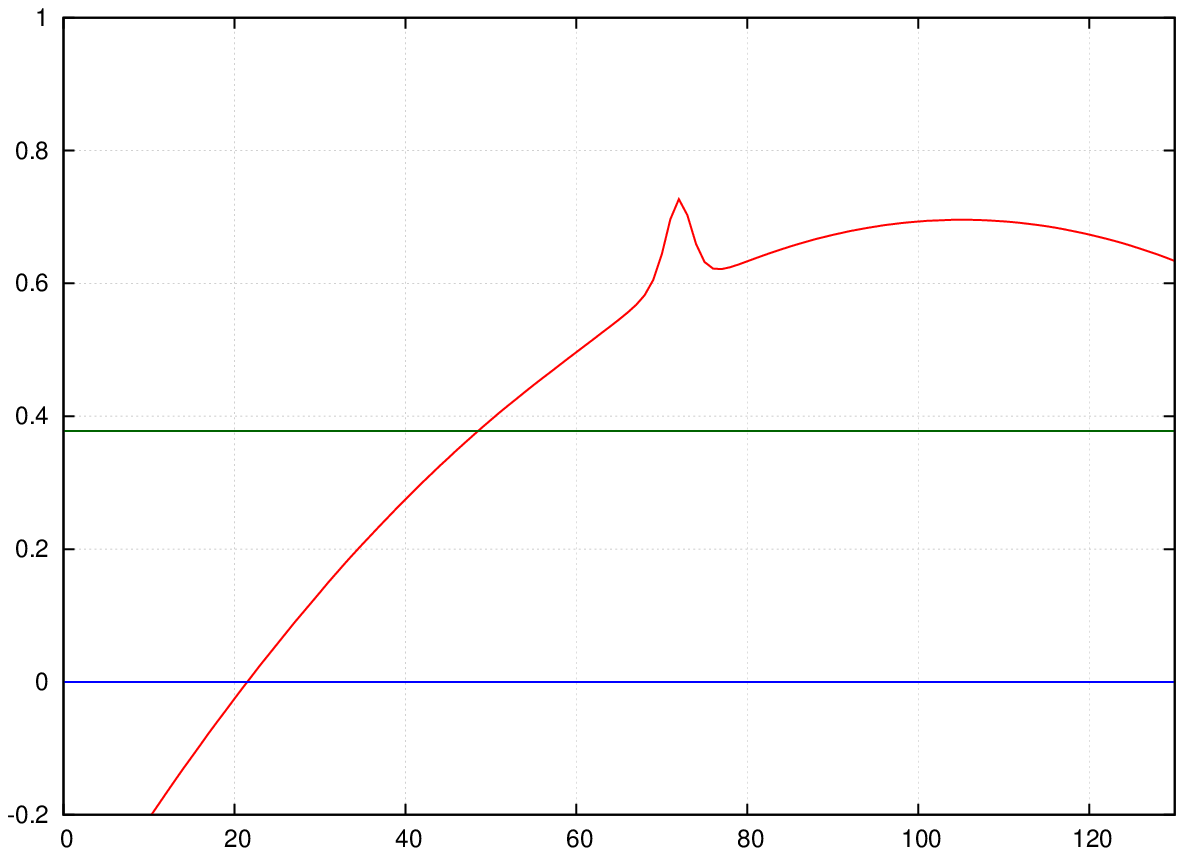} &
\includegraphics[width=0.5\linewidth]{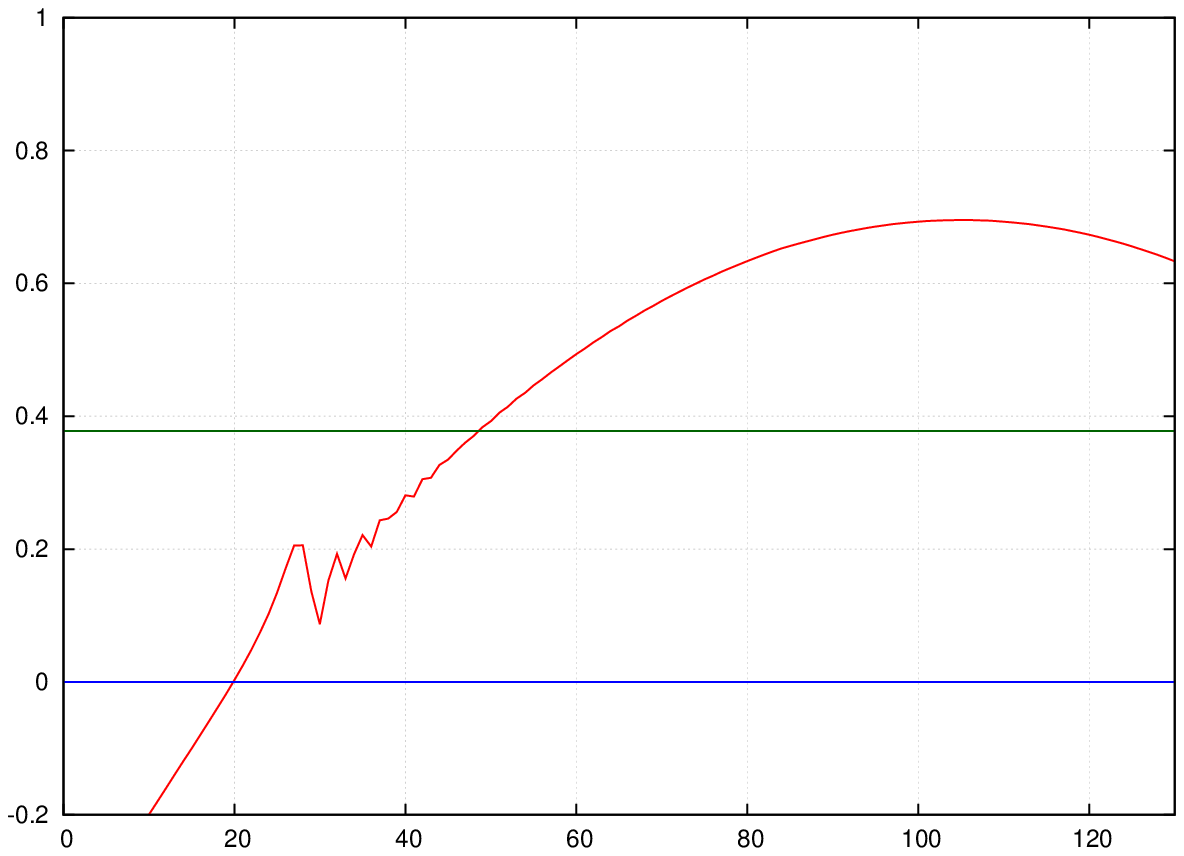}
\end{tabular}
\caption{Soliton dynamics in the presence of parabolic potential.
Soliton propagation is stable when parabolic potential is not
taking the system across boundaries of different regions(figs a-c).
Crossing the phase boundary
causes the soliton break-up (fig d). Green line marks the boundary
between regions I$_+$
and II$_+$.
}
\label{ParabolicDensity}

\end{figure}

{\it Experimental considerations}. Direct experimental tests of theoretical results presented in this paper
can be obtained by analyzing dynamics starting from soliton configurations
of ultracold atoms in optical lattices, such as shown in figs
\ref{SolitonPlusMinus}, \ref{TwoDimMinus}, \ref{TwoDimPlus}. These figures show that lattices of the
order of tens of sites should be sufficient for such studies.
Experimentally soliton like initial states can be created and their dynamics
can be studied
using single site resolution and addressability available in current experiments
with ultracold atoms in optical lattices.
Density modulation can be achieved by letting the system reach equilibrium
with a specified inhomogeneous potential. Phase modulation can be done by applying a short potential
pulse. We note that clear signatures of soliton dynamics can be achieved by starting
with initial states that have density modulation and no phase imprinting.
In this case a pair of solitons appears and propagates in the opposite directions.
This process takes place on relatively short time scales accessible to experiments.
Additional details about this procedure are discussed in the supplementary material.
We also note that to observe soliton dynamics one does not need to create initial configuration,
which match theoretically calculated solitonic solutions exactly.
Numerical analysis of (\ref{xyzSystem}) shows that initial conditions with spatially
localized  excess or deficit of magnetization (density) separate reasonably fast
into solitons and the wave background. On the other hand observing all stages of the magnetization (density) step
decay requires longer times and larger system sizes. Another important experimental consideration, which we did not address so far,
is the presence of the parabolic confining potential in all of the currently available experimental set-ups.
In fig \ref{ParabolicDensity} we show that as long as the chemical potential change is not taking the system across
one of the boundaries in fig. \ref{PhaseDiagram},  it has no strong effect on the soliton propagation.
However crossing any of the boundaries leads to the soliton break-up.

{\it Summary}. We analyzed semiclassical dynamics of anisotropic Heisenberg models in
dimension higher than one. Combining analytical study of small amplitude, long wavelength excitations with numerical studies of large amplitude, short wavelength excitations
we demonstrated the existence of a universal dynamical phase diagram, in which different
regions can be distinguished
based on the character of one dimensional solitonic excitations, and
stability of such solitons to transverse modulation.
Universality of
dynamics, which we find from direct solution of lattice equations, is very
intriguing.  Our model is not a special lattice regularization
of an integrable continuum system. We analyze lattice model describing
real physical systems which, in principle, can contain many terms
breaking universality.

We thank  I. Bloch, M. Greiner, M. Lewenstein, D. Pekker, and L. Pitaevskii for
insightful discussions. This work was partially supported  by the NSF Grant No. DMR-07-05472,
DARPA OLE program, CUA, AFOSR Quantum Simulation MURI, AFOSR MURI on Ultracold Molecules,
the ARO-MURI on Atomtronics (E.D.), Russian Federation Government Grant No. 2010-220-01-077,
Grant RFBR No. 11-01-12067-ofi-m-2011 (A.M.).
We also acknowledge support from the Harvard ITAMP and the Russian
Quantum Center (RQC).



\begin{center}
{\bf Supplementary material}
\end{center}

{\it Semiclassical dynamics}. We consider time-dependent variational
wavefunctions

\begin{equation}
\label{WaveFunction}
|\Psi (t) \rangle  =  \prod_{i} \left[ \sin {\theta_{i} (t)\over 2}
\, e^{-i \varphi_{i}(t) \over 2}  |\downarrow \rangle_{i}  +
\cos {\theta_{i}(t) \over 2}  \, e^{i \varphi_{i}(t) \over 2} \,
|\uparrow \rangle_{i} \right]
\end{equation}
To project quantum dynamics into these wavefunctions we define the Lagrangian
\cite{Jackiw1979,Huber2008,Demler2011a}
\begin{equation}
\label{AvLagr}
\begin{array}{c}
{\cal L} \, = \, - i \, \langle \Psi | \frac{ d}{ dt} | \Psi \rangle
+  \langle \Psi | H | \Psi \rangle \,\, =
\cr
\cr
= \, \sum_{i} {1 \over 2} \, {\dot \varphi}_{i} \,
\cos \theta_{i} \, - \, J_{z} \, \sum_{i, j=i\pm1}
\cos \theta_{i} \, \cos \theta_{j} \, -
\cr
\cr
- \, J_{\perp} \, \sum_{i, j=i\pm1} \sin \theta_{i} \, \sin \theta_{j} \,
\cos (\varphi_{i} - \varphi_{j})
\end{array}
\end{equation}
and write Lagrange equations for all variables. We obtain
\begin{equation}
\label{DiscreteSystem}
\begin{array}{c}
{\dot \varphi}_{i}  \sin  \theta_{i}  =
4  J_{z}  \sin  \theta_{i}  \left( \cos \theta_{i+1}
+  \cos \theta_{i-1} \right) -
\cr
- 4 J_{\perp} \cos \theta_{i} \left( \sin \theta_{i+1}
\cos (\varphi_{i} - \varphi_{i+1}) + \sin \theta_{i-1}
\cos (\varphi_{i} - \varphi_{i-1}) \right)
\cr
\cr
{\dot \theta}_{i} = -  4 J_{\perp} \left(
\sin \theta_{i+1} \sin (\varphi_{i} - \varphi_{i+1}) +
\sin \theta_{i-1} \sin (\varphi_{i} - \varphi_{i-1}) \right)
\end{array}
\end{equation} Introducing variables $x_i$, $y_i$ and $z_i$
we obtain equations (2) of the main text.

{\it Nonlinear Hydrodynamics}.
To describe the regime of long wavelength
modulations we introduce slow variables in space and time
$X = h x$, $T = h t$, where $h$ is the lattice constant.
We also define $\mu = \cos \theta$ and
$\sigma (X, T) =  h \varphi (X, t)$. Taking $h$
to be a small parameter we obtain
\begin{equation}
\begin{array}{c}
\label{ApprLagr}
{\cal L} \,\, = \,\, {1 \over 2} \, \sigma_{T} \, \mu \, - \,
2 \, J_{\perp} \, (1 - \mu^{2}) \, \cos \sigma_{X} \, - \,
2 \, J_{z} \, \mu^{2} \, +
\cr
\cr
+ \, h^{2} \, J_{\perp} \, (1 - \mu^{2}) \,
\left( {1 \over 3} \, \sigma_{XXX} \, \sin \sigma_{X} \, + \,
{1 \over 4} \, \sigma_{XX}^{2} \, \cos \sigma_{X} \right)
\, +
\cr
\cr
+ \, h^{2} \, J_{\perp} \,
{\mu^{2} \mu_{X}^{2} \over 1 - \mu^{2}} \, \cos \sigma_{X}
\, - \,  h^{2} \, J_{z} \, \mu \, \mu_{XX} \,\, + \,\,
{\cal O} (h^{4})
\end{array}
\end{equation}
Linearizing equations of motion around a uniform state we have
\begin{equation}
\label{LambdaSyst}
r^{1,2}_{T} \, = \,\pm \,  \lambda (r^{1}, r^{2}) \, r^{1,2}_{X}
\end{equation}
with
$
\lambda \, = \, 4 \, \sqrt{2 \, J_{\perp} \, (1 - \mu^{2}) \,
(\, J_{\perp} \,  -  \, J_{z})}
$.
Here $r^{1,2}$ are Riemann invariants \cite{Demler2011a}
that have physical interpretation of left and right moving
perturbations. When $\lambda$ is real we have hyperbolic regime
of the system, when $\lambda$ is imaginary we have elliptic regime.
Special lines $J_z/J_\perp=1$ and $\mu_0 = \pm 1$ have $\lambda=0$
and correspond to the parabolic regime.

{\it Relation to Korteweg–de Vries equations and solitons}.
To understand the character of solitons in the hyperbolic regime
we take equations of motion obtained from (\ref{ApprLagr}) and
keep only the lowest order terms in nonlinearity.
We further assume that left and right moving components of the perturbation
can be considered separately (e.g. we assume that left/right moving parts
separate spatially before effects of nonlinearity become important)
\begin{equation}
\label{r1disp}
r^{1}_{T} \,\, = \,\, - \, 6 \, \mu_{0} \,
\sqrt{J_{\perp} (J_{\perp} - J_{z})} \,\, r^{1} \,\, r^{1}_{X} \, +
\end{equation}
$$+ \, h^{2}
\sqrt{{2 J_{\perp} (1 - \mu_{0}^{2}) \over J_{\perp} - J_{z}}}
\left( J_{\perp} \left( {1 \over 6}  -
{\mu_{0}^{2} \over 1 - \mu_{0}^{2}} \right)  -
{7 \over 6} J_{z} \right) r^{1}_{XXX} $$

\begin{equation}
\label{r2disp}
r^{2}_{T} \,\, = \,\, 6 \, \mu_{0} \,
\sqrt{J_{\perp} (J_{\perp} - J_{z})} \,\, r^{2} \,\, r^{2}_{X} \, -
\end{equation}
$$-  \, h^{2}
\sqrt{{2 J_{\perp} (1 - \mu_{0}^{2}) \over J_{\perp} - J_{z}}}
\left( J_{\perp} \left( {1 \over 6}  -
{\mu_{0}^{2} \over 1 - \mu_{0}^{2}} \right)  -
{7 \over 6} J_{z} \right) r^{2}_{XXX} $$
Equations (\ref{r1disp}),
(\ref{r2disp}) are of the KdV type except for special lines
\begin{equation}
\label{boundary}
\mu_{0} \, = \, 0 \,\,\,\,\,\,\,\, , \,\,\,\,\,\,\,\,
J_{\perp} \, \left( {1 \over 6} \, - \,
{\mu_{0}^{2} \over 1 - \mu_{0}^{2}} \right) \, - \,
{7 \over 6} \, J_{z} \, \neq \, 0
\end{equation}
Analysis of solitons, including their stability to transverse modulation
can now be performed as discussed in ref. \cite{Demler2011a}.
Equation (\ref{boundary}) gives phase boundaries shown in fig. 1 of the main text.

{\it Imperfect preparation of the initial state}.

\begin{figure}

\begin{tabular}{cc}
\includegraphics[width=0.5\linewidth]{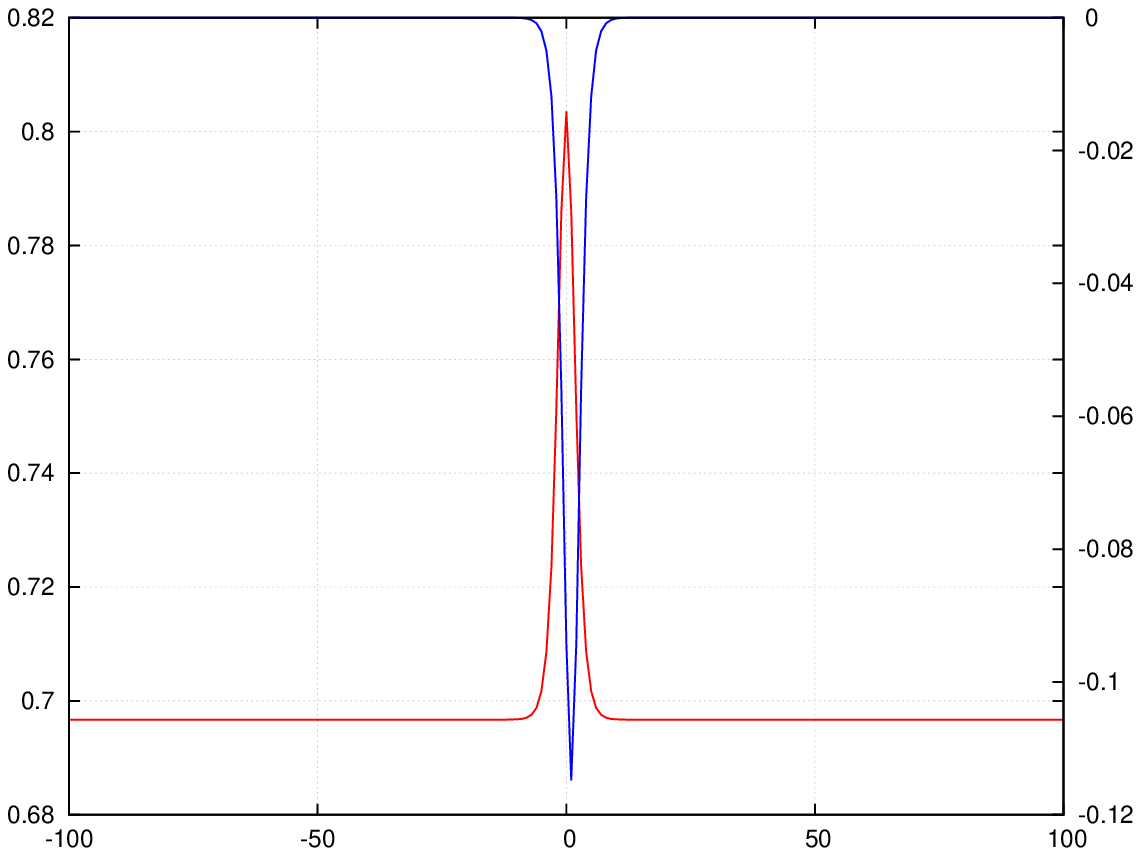} &
\includegraphics[width=0.5\linewidth]{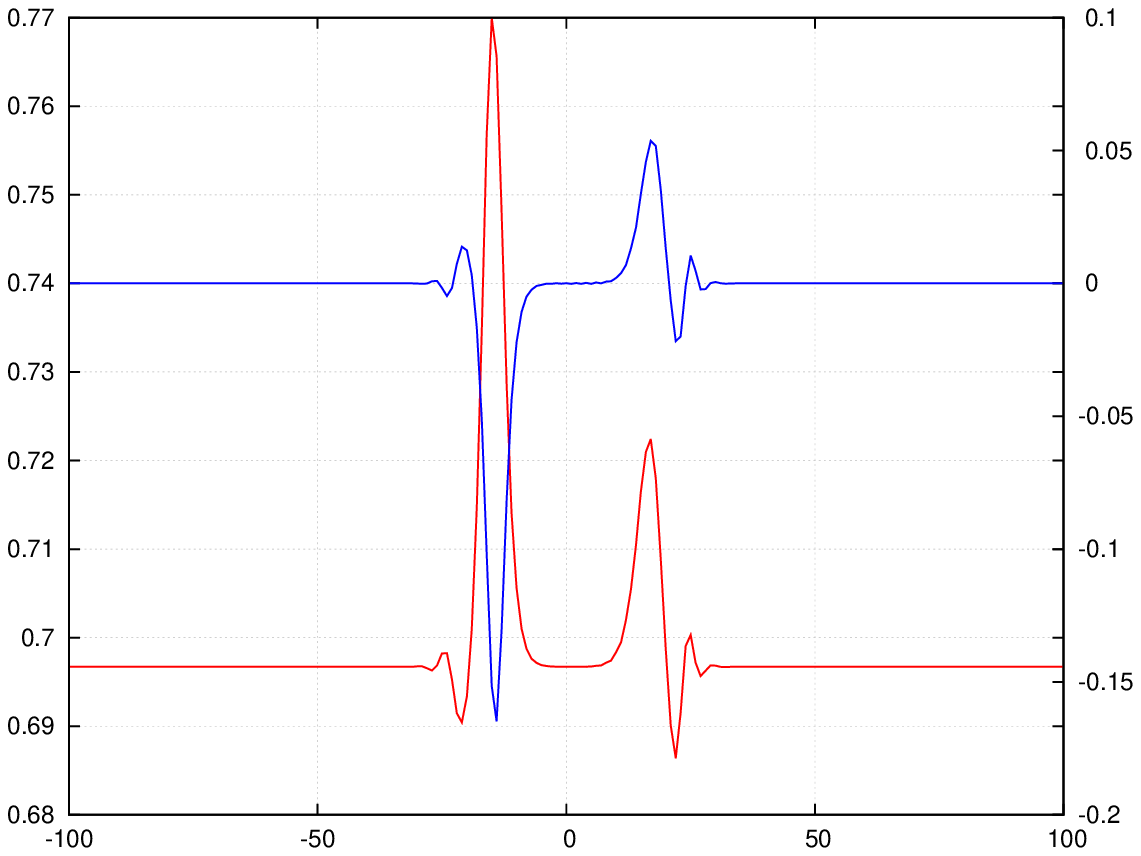} \cr
\includegraphics[width=0.5\linewidth]{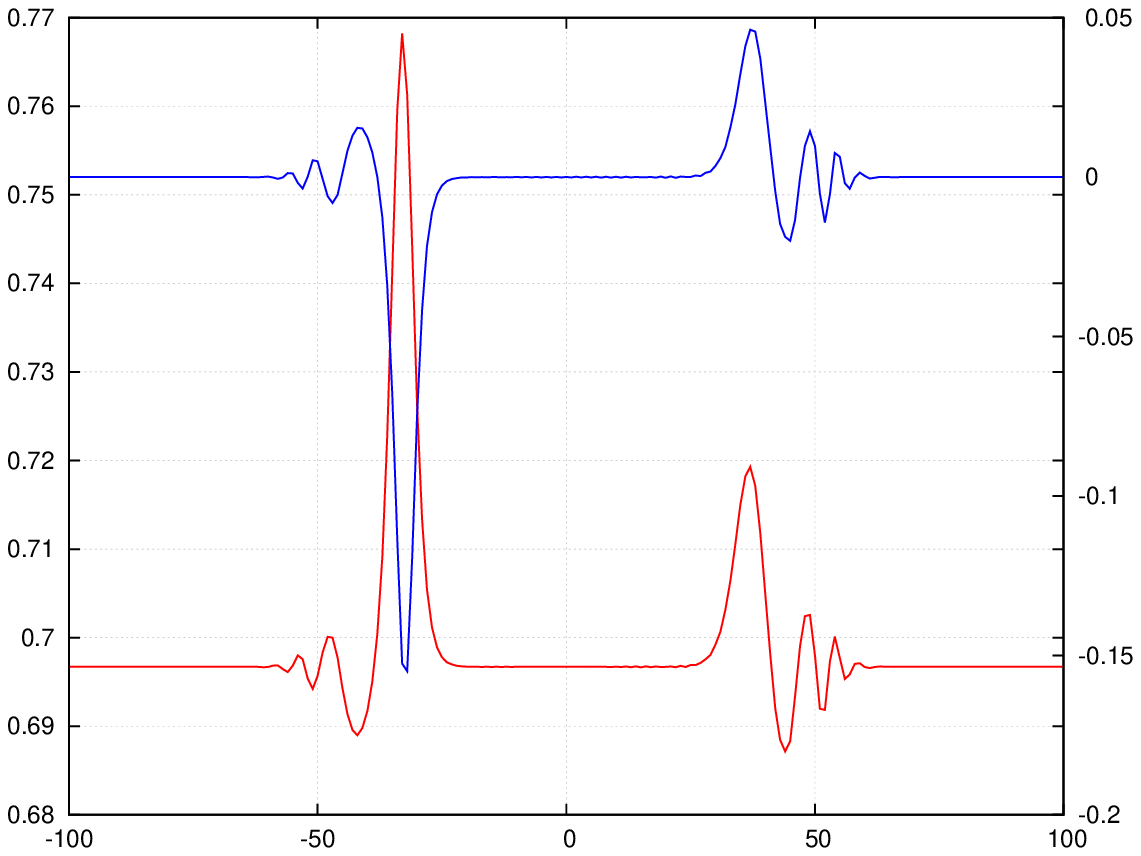} &
\includegraphics[width=0.5\linewidth]{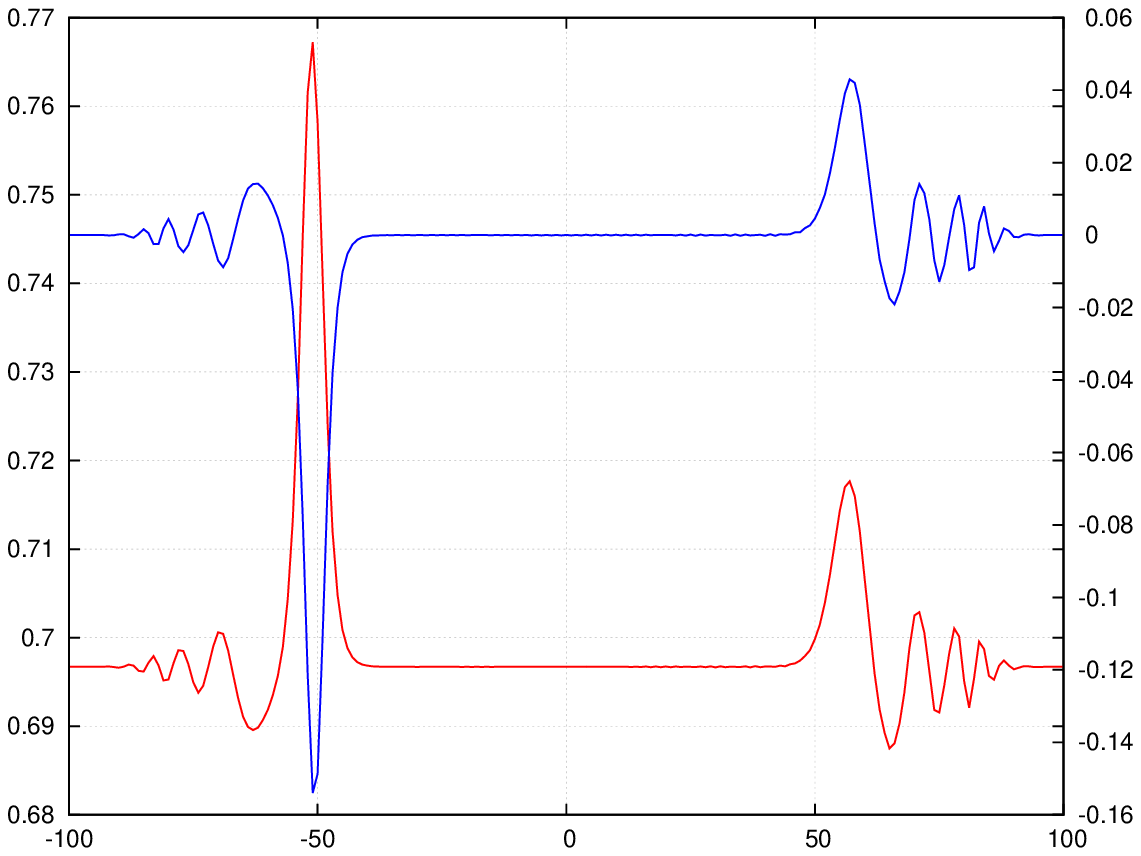}
\end{tabular}
\caption{Dynamics from a general initial state, which does not match a solitonic solution.
Red lines show magnetization (density), blue lines show superfluid velocity,
$\phi_i - \phi_{i+1}$. Initial parameters
correspond to region ${\rm I}_{+}$.
Different plots are for $t J_\perp = 0$, $5$, $10$, $15$.
Note separation
of perturbations into solitons and wave-trains.
}
\label{SolitonPlusSep}

\end{figure}

An important question for experimental
realizations of solitons  is dynamics from initial
states that do not match exactly soliton solutions. Fig. \ref{SolitonPlusSep} shows that in this case the system
separates perturbations into solitons and
wave-trains.

{\it Parabolic regime}.

\begin{figure}

\begin{tabular}{cc}
\includegraphics[width=0.45\linewidth]{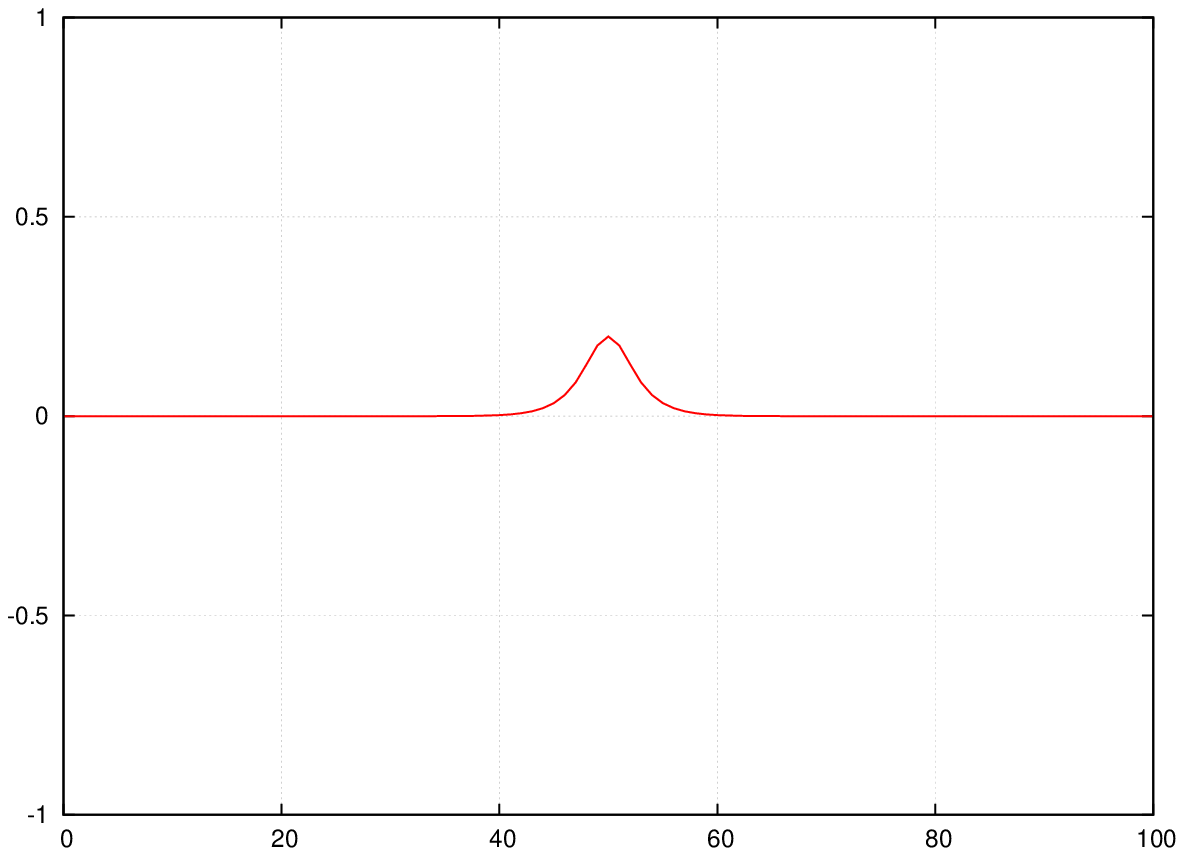} &
\includegraphics[width=0.45\linewidth]{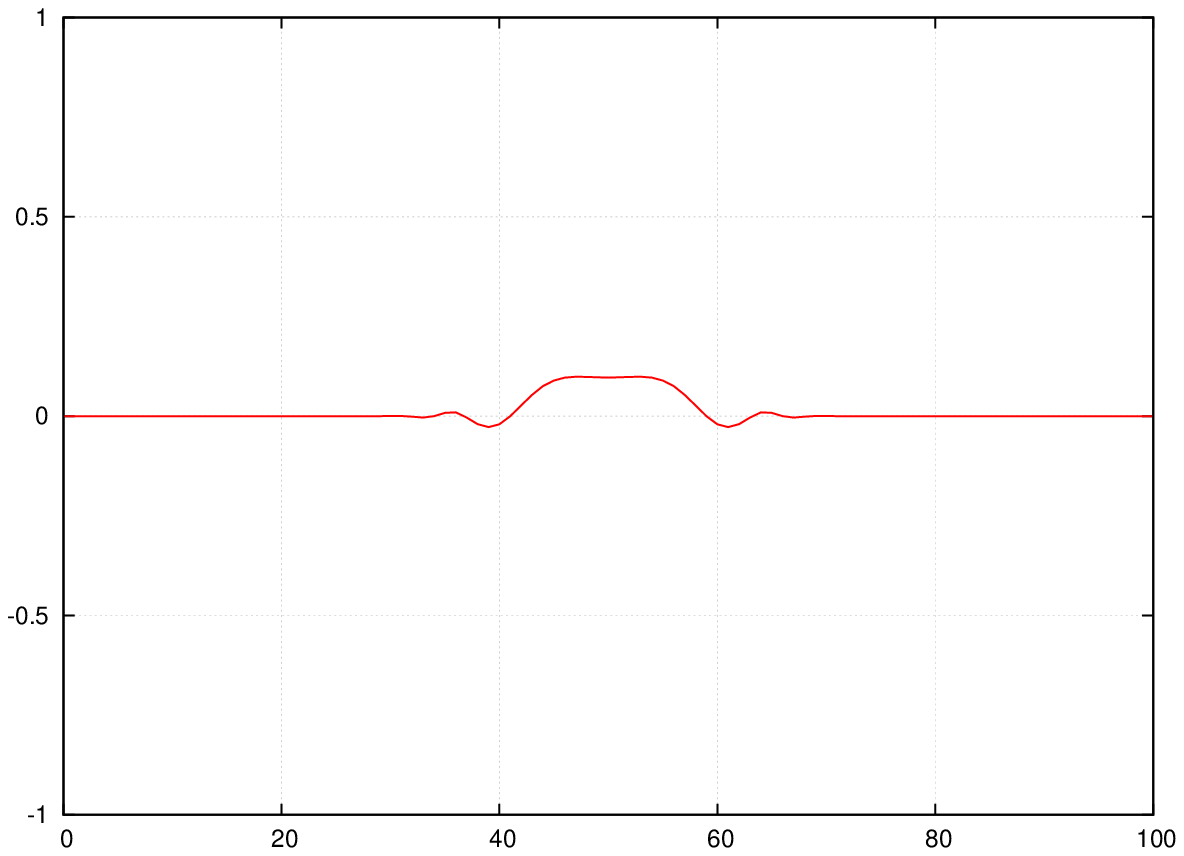} \cr
\includegraphics[width=0.45\linewidth]{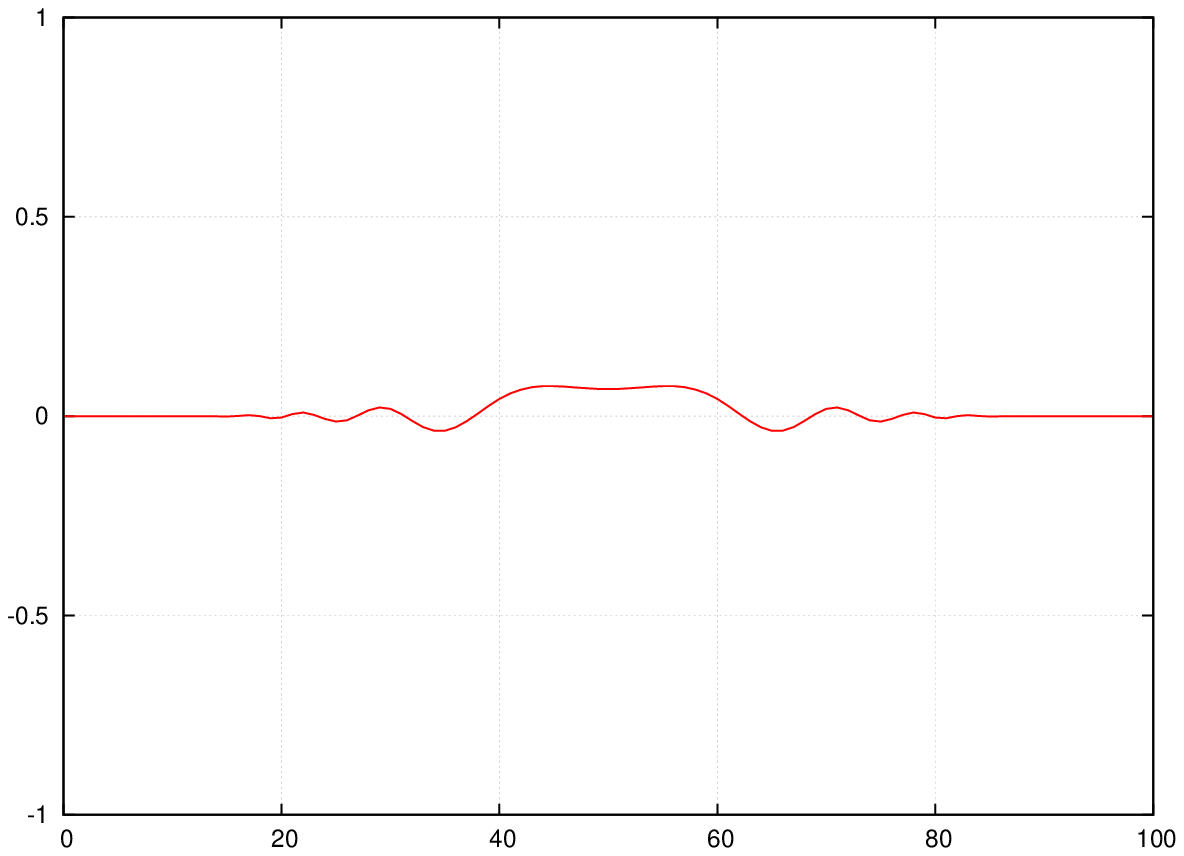} &
\includegraphics[width=0.45\linewidth]{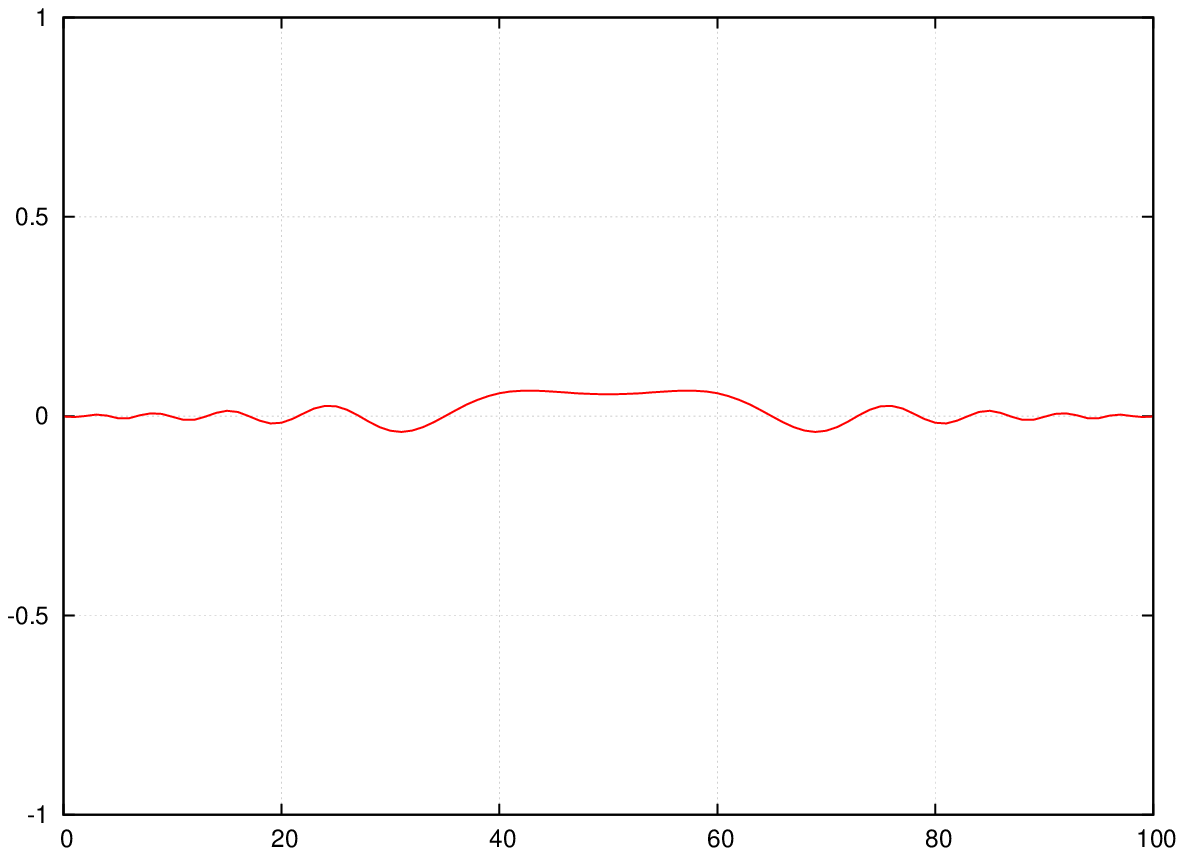}
\end{tabular}
\caption{Time evolution of a small density
perturbation on the parabolic line.
Vertical axis is magnetization (density).
Note a slowly growing oscillation region.
Different plots correspond to $t J_\perp = 0$, $2$, $4$,
$6$.}
\label{Parabolic}

\end{figure}

In Fig. \ref{Parabolic} we show dynamics of solutions
in the parabolic regime separating the elliptic and hyperbolic regions of the phase diagram 1 in the main text.

\end{document}